\documentclass[11pt]{article}

\usepackage[utf8]{inputenc}  
\usepackage[T1]{fontenc}
\usepackage{graphicx}
\usepackage[table]{xcolor}% http://ctan.org/pkg/xcolor
\usepackage{amsmath}

\usepackage{lmodern}
\usepackage{microtype}
\usepackage{amsmath,amssymb}%,amsfonts,amssymb,amsthm
\usepackage{mathrsfs}
\usepackage[table]{xcolor} 
\usepackage{graphicx}
\usepackage{braket}
\usepackage{mleftright} %\mleftright
\usepackage{array}
\usepackage{booktabs}
\usepackage{slashed}
\usepackage{color}   

\usepackage[english]{babel} 
\usepackage{caption} 
\usepackage{subcaption} %for subfigures
\usepackage{bm} %for making symbols bold by \bm command   
\usepackage{multirow} %for drawing table
\usepackage{calc} 
\usepackage{makecell} %for adjusting table /hline thickness
\usepackage{physics}
\usepackage{url}
\usepackage{hyperref}
\usepackage{cleveref}
\usepackage{ragged2e}     % command \RaggedRight
\usepackage{comment}

\newcommand{\ra}{ 
\rightarrow
}

\mathchardef\mhyphen="2D %small hyphen in math mode

\definecolor{blizzardblue}{rgb}{0.93, 0.93, 0.93} % light gray
\newcolumntype{?}{!{\vrule width 0.8pt}} %makes vertical lines in tables with thickness 0.8pt
\mathchardef\mhyphen="2D %small hyphen in math mode

\newcommand{\RN}[1]{ %command for writing roman numbers 
  \textup{\uppercase\expandafter{\romannumeral#1}}%  
\renewcommand\thesubfigure{(\alph{subfigure})} %for referencing to subfigure as 2(a) instead of 2a
\captionsetup[sub]{labelformat=simple} 
}

\begin{document}

\begin{center}

\vspace*{15mm}
\vspace{1cm}
{\Large \bf Concurrent Exploration of Axion-Like Particle Interactions with Gauge Bosons at the LHC}

\vspace{1cm}

\renewcommand{\thefootnote}{\fnsymbol{footnote}}
{\bf Shirin Chenarani$^{1,2}$}
and 
{\bf Mojtaba Mohammadi Najafabadi$^{1,3}$}

 \vspace*{0.5cm}
 
{\small\sl 
$^{1}$ School of Particles and Accelerators, Institute for Research in Fundamental Sciences (IPM), \\P.O. Box 19395-5531, Tehran, Iran}

{\small\sl 
$^{2}$ Department of Physics, University of Science and Technology of Mazandaran, \\P.O.Box 48518-78195, Behshahr, Iran}

{\small\sl 
$^{3}$Experimental Physics Department, CERN, 1211 Geneva 23, Switzerland }

\vspace*{.2cm}
\end{center}

\vspace*{5mm}

%***********************************************************
\begin{abstract}\label{abstract}

Axion-like particles (ALPs) are pseudo Nambu-Goldstone bosons associated with spontaneously broken 
global symmetries incorporated in the Standard Model (SM) Lagrangian in many models beyond the SM.
The existence of a light ALP is plausible due to the long-standing problems that the SM has not been able 
to address, such as the dark matter (DM) problem and the observed matter-antimatter asymmetry.
There are many proposals in recent decades considering the ALP as a solution to some of these shortcomings.
Motivated by such potential, we search for ALPs with a mass of 1 MeV at the LHC in a model-independent fashion.
We explore two complementary production modes: ALP production in association with a pair of electroweak gauge
bosons ($ZZ$ or $WW$) and ALP production in association with a single gauge boson ($W$ or $Z$) plus jets.
For the $VV+a$ final state, signal and dominant SM backgrounds are generated, and a realistic detector response simulation is performed.
A multivariate analysis is employed to discriminate the $VV+a$ signal from background processes, and the expected 
$95\%$ confidence level (CL) exclusion limits in two-dimensional parameter spaces involving the ALP couplings are subsequently derived.
The $V$+jets channel is interpreted using LHC measurements in regimes where the
 ALP escapes detection, appearing as missing energy.
 The two analyses are complementary: both the $VV+a$ and the $V$+jets channels probe simultaneously the 
 ALP couplings to gluons and to electroweak gauge bosons. While the $VV+a$ channel offers clean multi-lepton 
 final states and direct reconstruction of the $ZZ$ or $WW$ system, the $V$+jets channel benefits from larger 
 production cross sections, enabling stronger constraints in certain regions of the parameter space.
Comparing the limits obtained in this study with current and projected limits derived from LHC searches, 
it is seen that competitive sensitivities to the ALP couplings are achieved, and a significant region of previously unexplored parameter space becomes accessible.
\end{abstract}

\newpage

%***********************************************************
\section{Introduction}
\label{sec:introduction}
The Standard Model (SM) of particle physics is a successful theory with a large number of experimentally confirmed predictions. However, there are some problems that the SM is not able to address. For example, the dark matter (DM) problem, baryon asymmetry problem, strong CP problem, anomalous magnetic dipole moment of muon, etc. An extensive effort has been made to solve some of the SM shortcomings by developing models beyond the SM. One of the ideas that has attracted much attention is the existence of Axions which was originally proposed to address the strong CP problem \cite{Peccei:1977hh,Hook:2018dlk,Dine:2000cj,DiLuzio:2020wdo,Hook:2019qoh,Quevillon:2019zrd,Bellazzini:2017neg,Arganda:2018cuz, Dine:1981rt}. Axions are pseudo Nambu-Goldstone bosons associated with a spontaneously broken global $U(1)$ symmetry. The Axion mass is stringently related to the scale of the global symmetry breaking. Setting this relation aside, Axion-like particles (ALPs) with broad-ranging masses and couplings are introduced. ALPs exist in any model with a spontaneously broken global symmetry. They can be viable candidates for a DM particle \cite{Preskill:1982cy,Abbott:1982af,Dine:1982ah}, and can also explain the observed matter-antimatter asymmetry \cite{Jeong:2018jqe,Co:2019wyp}. In addition, ALPs provide the possibility to explain the anomalous magnetic dipole moment of muon \cite{Bauer:2019gfk,Cornella:2019uxs}, the anomalous decays of the excited Beryllium $^{8}\mathrm{Be}^*$  \cite{Krasznahorkay:2015iga,Ellwanger:2016wfe}, and the excess of events observed in the rare $K$ mesons searches performed by the KOTO experiment \cite{Kitahara:2019lws}. There are proposals that the neutrino mass problem may also be solved by ALPs through a coupling to neutrinos \cite{Dias:2014osa, Chen:2012baa, Salvio:2015cja}. Being motivated by such possibilities, the ALP parameter space has been extensively probed to date by collider searches, low-energy experiments, cosmological observations, etc. \cite{Brivio:2017ije,Biswas:2023ksj,Choi:2020rgn,Athron:2020maw,Han:2020dwo,Mimasu:2014nea,Bauer:2017ris,Aloni:2019ruo,Haghighat:2021djz,Haghighat:2020nuh,Haghighat:2022qyh,Ebadi:2019gij, Hosseini:2024kuh, Hosseini:2022tac, ARGUSlimit,Baldini:2020okg,Aad:2019ugc,Calibbi:2020jvd,Iguro:2020rby,Endo:2020mev, c144, c244, c344}.

Light ALPs with masses below the threshold of the electron pair production ($\approx 1$ MeV) predominantly decay into a pair of photons. Other decay channels, e.g. decays into leptons and hadrons, become available for heavier ALPs. ALPs with large enough masses decay promptly after production and can be reconstructed by their decay products. Many studies have searched for such ALPs. The exotic Higgs decays $h\rightarrow aZ$ and $h\rightarrow aa$ followed by the ALP decays $a\rightarrow\gamma\gamma$ and $a\rightarrow\ell\ell$ have been studied at the LHC \cite{Bauer:2017ris,Bauer:2017nlg,Chatrchyan:2012cg,Khachatryan:2017mnf}. The tri-$\gamma$ production, $\gamma a\ra 3\gamma$, and the associated production of an ALP and a Higgs boson (or a top quark pair) have been studied in Refs. \cite{Brivio:2017ije,Mimasu:2014nea,Bauer:2017ris,Jaeckel:2015jla}. The $Z\rightarrow \gamma\gamma$ data from LEP has been analyzed providing significant sensitivities to the ALP mass range MeV to 90 GeV \cite{Jaeckel:2015jla}. There are also a number of studies that utilizes the ultra-peripheral heavy-ion collisions at the LHC to search for ALPs with masses from 100 MeV to 100 GeV that decay into photons promptly \cite{Knapen:2016moh}. 

The width of the ALP decay into a photon pair scales as the third power of the ALP mass, $m_a^3$. Moreover, the ALP coupling to photons is strongly constrained by observations \cite{Bauer:2017ris}. Therefore, for light ALPs with masses $\lesssim 1$ MeV, which predominantly decay into photons, the decay length becomes so large. This motivates helioscope experiments such as SUMICO \cite{Inoue:2008zp}, CAST \cite{Arik:2008mq} and IAXO  \cite{Irastorza:2013dav} which search for solar ALPs. Proton beam-dump experiments have also been performed to probe long-lived ALPs with masses in the MeV-GeV range \cite{Dobrich:2019dxc}. In addition, collider studies have been performed in search of long-lived ALPs. Mono-$\gamma$ and mono-jet plus missing energy have been shown to be promising signals to probe ALP masses above the KeV scale \cite{Mimasu:2014nea}. Mono-$W$, mono-$Z$, $W\gamma$, $Z\gamma$, $WW$ and $WW\gamma$ plus missing energy signals have been studied at the LHC to probe the ALP mass and couplings to electroweak gauge bosons \cite{Brivio:2017ije,Biswas:2023ksj,ATLAS:2018sxj,CMS:2019ppl,CMS:2023rcv,ATLAS:2019lsy,CMS:2017ret,TheATLAScollaboration:2015vog,CMS:2017nxf}. 
Other experimental searches for ALPs have been performed by ATLAS and CMS, including in anomalous Higgs decays to photon pairs \cite{c1}, 
light-by-light scattering in heavy-ion collisions \cite{c2}, and nonresonant diboson production such as $ZZ$ and $ZH$ final states \cite{c3}. 
These studies place direct constraints on ALP couplings to photons, gluons, and electroweak gauge bosons across a broad mass range.

Although much effort has been devoted to probing light ALPs, a significant region of their parameter space remains unexplored. 
Such ALPs offer the potential to address persistent issues in the SM, such as the dark matter (DM) 
problem \cite{Preskill:1982cy,Abbott:1982af,Dine:1982ah}, the baryon asymmetry problem \cite{Jeong:2018jqe,Co:2019wyp}, 
and the anomalous magnetic dipole moment of the muon \cite{Bauer:2019gfk,Cornella:2019uxs}. 
Motivated by these possibilities, we conduct a collider search for long-lived ALPs with a mass of 1 MeV in this work. 
These ALPs can be produced at colliders through their couplings to SM particles. 
In the first part of this work, we assume the collider to be the High Luminosity Large Hadron Collider (HL-LHC) \cite{ZurbanoFernandez:2020cco} 
operating at a center-of-mass energy of 14 TeV. 
Given that the assumed ALP mass is much smaller than the characteristic energy of the interactions, 
the study's results would remain largely unaffected by considering even lighter ALP masses. 
ALPs can be produced in association with single or pairs of electroweak gauge bosons, such as $ZZ$ or $WW$, at the LHC. 
In this work, we focus on simultaneously probing ALP couplings with gluons and $W$ bosons. 
Probing multiple couplings concurrently provides a more comprehensive understanding of the ALP parameter 
space and can reveal interactions that might be missed when considering individual couplings in isolation. 
This approach allows for more stringent and realistic constraints on ALP properties, thereby improving the 
sensitivity of our search and significantly expanding the explored parameter space.
Ref.\cite{ccvv} presents concurrent bounds on ALP couplings to electroweak gauge bosons, 
derived from Vector Boson Scattering (VBS) processes at the LHC.
The analysis focuses on non-resonant ALP-mediated VBS, where the ALP 
 serves as an off-shell mediator. This process becomes relevant when the ALP is too light to 
 be produced resonantly, exploiting the derivative nature of ALP interactions with the electroweak bosons of the SM. 
 The study investigates the production of vector boson (VV) pairs with large invariant masses, accompanied by two jets. 
 Operating within a gauge-invariant framework, the analysis extracts upper limits on ALP couplings to electroweak bosons 
 through a reinterpretation of public CMS VBS results from Run 2.

The focus of this study is on the long-lived ALPs tend to be stable at the detector and manifest themselves as missing energy. 
Searching for the missing energy signature, we constrain the ALP couplings to the SM particles, 
namely couplings to $W$ bosons and gluons. We provide several sets of limits obtained independently 
in the search for $ZZa$ and $WWa$ signals. The $ZZa$ analysis uses both the fully leptonic and semi-leptonic final states, 
and in the $aWW$ analysis, only the fully leptonic final state is considered. 
The analyses use a realistic detector response simulation and deploy a multivariate technique to discriminate the signal from background.
We provide prospects for the HL-LHC with the integrated luminosity of 3 $\mathrm{ab^{-1}}$ as well as the expected limits with 
Run 2 LHC data which corresponds to an integrated luminosity of 138 fb$^{-1}$. 
Additionally, we probe the parameter space of the ALP using interpretations of the LHC experimental 
data from $W$+jets and $Z$+jets measurements. These production processes are sensitive to ALP couplings to $W$ bosons, gluons, and fermions. 
As a result, the limits in this study are obtained assuming two non-zero ALP coupling constants at a time and are provided in two-dimensional planes. 
This approach enables us to provide concurrent limits on the ALP couplings, rather than on a single coupling. 
Comparing the limits obtained in this study with the current and projected limits inferred from the LHC searches, 
it will be shown that the present study improves the limits significantly and can be used to probe a vast region of unprobed parameter space.

It is notable that while the $VV+a$ channel directly probes interactions involving 
two electroweak gauge bosons and an ALP, the $V$+jets channel (with $V = Z, W^\pm$) 
benefits from higher production cross sections and more inclusive event selection criteria, 
enabling stronger constraints in certain regions of the parameter space. 
These two channels are therefore complementary in their sensitivity to different coupling structures and kinematic regimes. 
By concurrently investigating both channels within the same framework, we provide a more comprehensive and robust 
exploration of ALP-gauge boson interactions. 
This dual approach allows us to set simultaneous constraints on multiple ALP 
couplings and offers a clearer understanding of their interplay at the LHC.

This paper is structured as follows. In Section \ref{sec:lagrangian}, the effective Lagrangian describing an ALP and its interactions with the SM particles is introduced. In section \ref{sec:signalandbkg}, the signal processes under consideration and the relevant SM background processes are discussed. In section \ref{sec:analysis}, analysis details are discussed. In particular, section \ref{sec:eventgeneration} describes the event generation method and tools, section \ref{subsec:validity} discusses the validity of the effective Lagrangian, section \ref{selection} discusses the object identification and event selection, and section \ref{discrimination} provides the details of the signal-background discrimination. The method used to calculate the limits and the obtained results are provided in section \ref{sec:constraint}.
Section \ref{wjzj} focuses on interpreting the CMS experiment measurements of the differential cross sections for 
the $W+$jets and $Z+$jets processes to investigate the couplings of the ALP with gluons and $W$ bosons. 
Finally, the summary and conclusions of the paper are presented in Section \ref{sumconvvv}.

%***********************************************************
\section{ALPs Effective Lagrangian}  
\label{sec:lagrangian}
The QCD axion is the pseudo Nambu-Goldstone boson associated with the spontaneously broken anomalous global $U\mathrm{(1)}_{\mathrm PQ}$ symmetry incorporated in the SM Lagrangian in the model proposed by Peccei and Quinn \cite{Peccei:1977hh}. Providing the possibility of resolving the strong CP problem is the main motivation behind proposing the existence of the QCD axion. There is a strict relation between the mass and couplings of the QCD axion. ALPs are hypothetical particles resulting from eluding the relation connecting the mass and couplings of the QCD axion. Neglecting ALP couplings to fermions, the most general effective Lagrangian up to dimension-5 operators describing the interaction between the ALP and the SM fields is given by~\cite{Brivio:2017ije}
\begin{eqnarray}
\begin{aligned}
	\mathscr{L}_{\mathrm{eff}} &= \mathscr{L}_{\mathrm{SM}} + \frac{1}{2} (\partial^\mu a)(\partial_\mu a) - \frac{1}{2} m_{a,0}^2 a^2\\
	&+ c_{a \Phi} \frac{i \partial^\mu a}{f_a} (\Phi^\dagger \overleftrightarrow{D}_\mu \Phi)
	-  c_{BB} \frac{a}{f_a} B_{\mu\nu} \tilde{B}^{\mu\nu}
	-  c_{WW} \frac{a}{f_a} W^{A}_{\mu\nu} \tilde{W}^{\mu\nu,A}
	-  c_{GG} \frac{a}{f_a} G^{A}_{\mu\nu} \tilde{G}^{\mu\nu,A},
\end{aligned}
\label{lag}
\end{eqnarray} 
where $G^A_{\mu\nu}$, $W^A_{\mu\nu}$ and $B_{\mu\nu}$ respectively represent the gauge field strength tensors corresponding to $SU\mathrm{(3)}_c$, $SU\mathrm{(2)}_L$ and $U\mathrm{(1)}_Y$ symmetries, the dual field strength tensors $\tilde{X}^{\mu\nu}$ are defined as $\tilde{X}^{\mu\nu} \equiv \frac{1}{2} \epsilon^{\mu \nu \rho \sigma} X_{\rho \sigma}$ with $\epsilon^{\mu \nu \rho \sigma}$ being the Levi-Civita symbol, $f_a$ denotes the scale associated with the breakdown of the global $U\mathrm{(1)}$ symmetry, $a$ and $\Phi$ are respectively the ALP field and the Higgs boson doublet, and $\Phi^\dagger \overleftrightarrow{D}_\mu \Phi \equiv \Phi^\dagger D_\mu \Phi - (D_\mu \Phi)^\dagger \Phi$. The ALP field undergoes the translation $a(x)\rightarrow a(x)+\alpha$, with $\alpha$ being a constant, under a $U\mathrm{(1)}$ transformation (the symmetry corresponding to the $a$ field translation is called the shift symmetry). As a result, a $U\mathrm{(1)}$ invariant effective Lagrangian should only contain the derivative of the ALP field. This is not the case here since the Lagrangian has a chiral anomaly induced by the $aG\tilde{G}$ operator (the last operator in $\mathscr{L}_{\mathrm{eff}}$, Eq.~\ref{lag}) which couples the ALP field to the gluon density $G\tilde{G}$. This operator induces a total divergence contribution to $\delta\mathscr{L}_{\mathrm{eff}}$ under the ALP field translation. This total divergence results in a non-zero change in the action due to the QCD vacuum structure and instanton effects. The shift symmetry is also broken (softly) by the ALP mass term present in $\mathscr{L}_{\mathrm{eff}}$. The ALP mass term makes ALP acquire more mass than is obtained from non-perturbative dynamics and adds a mass to the dynamically generated ALP mass, $m_a^2=m_{a,0}^2+m_{a,\,\textrm{dyn}}^2$. 
This leads to a large parameter space and thus a rich phenomenology for ALPs. 
After electroweak symmetry breaking (EWSB), the operator $c_{a \Phi} \frac{i \partial^\mu a}{f_a} (\Phi^\dagger \overleftrightarrow{D}_\mu \Phi)$
in  $\mathscr{L}_{\mathrm{eff}}$ gives rise to a term of the form $Z_\mu \partial^\mu a$, which complicates the direct interpretation of the effective Lagrangian.
A field redefinition of the form
\begin{equation}
\begin{aligned}
\Phi\to \, e^{i\alpha_{\Phi}\, a/f_a}\Phi,\,\,\,\,\,\,
\psi_L\to \, e^{i\alpha_{\psi L}\, a/f_a}\psi_L,\,\,\,\,\,\,
\psi_R\to \, e^{i\alpha_{\psi R}\, a/f_a}\psi_R,
\label{redef}
\end{aligned}
\end{equation}
where $\psi_L = \{Q_L, L_L\}$, $\psi_R = \{u_R, d_R, e_R\}$, $\alpha_{\Phi}$ is a real constant and $\alpha_{\psi L,R}$ are $3\times 3$ hermitian matrices in flavor space, is thus applied to interpret the Lagrangian. The $a$-$Z$ term can be redefined away in favour of the fermionic couplings of the ALP with the choice $\alpha_\Phi=c_{a \Phi}$. The structure of the fermionic couplings induced in this way (Yukawa-like, vector-axial or a combination of them) can be controlled by tuning the parameters $\alpha_{\psi L,R}$~\cite{Brivio:2017ije}. Applying the field redefinitions in Eq.~\ref{redef} with the assumptions $\alpha_\Phi = c_{a \Phi}$ and $\alpha_{\psi L,R}=0$ leads to \begin{eqnarray} 
\begin{aligned}
	\mathscr{L}_{\mathrm{SM}} \to \,\, \mathscr{L}_{\mathrm{SM}} &+ c_{a \Phi} \left[i\left(\bar{Q}_L \mathbf{Y}_U\tilde\Phi u_R-\bar{Q}_L \mathbf{Y}_D\Phi d_R-\bar{L}_L\mathbf{Y}_E\Phi e_R\right)\frac{a}{f_a} + \text{h.c.}\right]\\
	&- c_{a \Phi} \frac{i \partial^\mu a}{f_a} (\Phi^\dagger \overleftrightarrow{D}_\mu \Phi),
\end{aligned}
\label{smredef}
\end{eqnarray} 
where $\tilde\Phi = i\sigma^2\Phi^*$ and $\mathbf{Y}_U$, $\mathbf{Y}_D$ and $\mathbf{Y}_E$ are respectively $3\times3$ matrices in flavor space containing Yukawa coupling constants for up-type quarks, down-type quarks and charged leptons. The Yukawa-like ALP-fermion operators in the first line of Eq.~\ref{smredef}, which stem from the SM Yukawa Lagrangian, can be conveniently written as
\begin{eqnarray}  
	c_{a \Phi} \left[i \frac{a}{f_a} \sum_{\psi=Q,L} \left(\bar{\psi}_L \mathbf{Y}_\psi \mathbf{\Phi} \sigma_3 \psi_R\right) + \text{h.c.}\right],
\label{fermionicops}
\end{eqnarray} 
where $Q_R\equiv\{u_R,d_R\}$, $L_R\equiv\{0,e_R\}$, $\sigma_3$ acts on weak isospin space, and the block matrices $\mathbf{\Phi}$ and $\mathbf{Y}_\psi$ are given by
\begin{eqnarray}
	\mathbf{\Phi} = \textrm{diag}(\tilde\Phi, \Phi), \ \ \ \ 
	\mathbf{Y}_Q \equiv \textrm{diag}(\mathbf{Y}_U,\mathbf{Y}_D), \ \ \ \ 
	\mathbf{Y}_L \equiv \textrm{diag}(0,\mathbf{Y}_E).
\label{blockmatrices}
\end{eqnarray} 
The operator \( c_{a \Phi} \frac{i \partial^\mu a}{f_a} (\Phi^\dagger \overleftrightarrow{D}_\mu \Phi) \) in \( \mathscr{L}_{\mathrm{eff}} \) introduces a derivative coupling between the ALP and the Higgs field. However, this term can be eliminated by a field redefinition of the Higgs doublet, \( \Phi \rightarrow e^{i c_{a\Phi} a/f_a} \Phi \), which generates an operator with opposite sign that cancels the original term, as shown in Eq.~\ref{smredef}. Applying the field redefinitions, the effective Lagrangian up to dimension-5 operators becomes~\cite{Brivio:2017ije}:
\begin{eqnarray}
\begin{aligned}
	\mathscr{L}_{\mathrm{eff}} = 
	& \, \mathscr{L}_{\mathrm{SM}} 
	+ \frac{1}{2} (\partial^\mu a)(\partial_\mu a) 
	- \frac{1}{2} m_{a,0}^2 a^2 \\
	& + c_{a \Phi} \left[i \frac{a}{f_a} \sum_{\psi=Q,L} \left(\bar{\psi}_L  \mathbf{Y}_\psi \Phi \sigma_3 \psi_R\right) + \text{h.c.}\right] \\
   & - c_{BB} \frac{a}{f_a} B_{\mu\nu} \tilde{B}^{\mu\nu}
     - c_{WW} \frac{a}{f_a} W^{A}_{\mu\nu} \tilde{W}^{\mu\nu,A}
	 - c_{GG} \frac{a}{f_a} G^{A}_{\mu\nu} \tilde{G}^{\mu\nu,A}.
\end{aligned}
\label{lagfinal}
\end{eqnarray}
In this expression,  \( \mathscr{L}_{\mathrm{SM}} \) is the SM Lagrangian, the second and third terms describe the canonical kinetic and mass terms for the ALP \( a \), 
and the fourth term encodes ALP-fermion interactions via Yukawa couplings after electroweak symmetry breaking (EWSB).
The remaining terms in $\mathscr{L}_{\mathrm{eff}}$ correspond to anomalous ALP couplings to the SM 
gauge bosons: hypercharge (\( B_{\mu\nu} \)), weak (\( W_{\mu\nu}^A \)), and gluon (\( G_{\mu\nu}^A \)) field strengths.
After EWSB, the neutral gauge bosons \( B_\mu \) and \( W_\mu^3 \) mix to form the photon and \( Z \) boson. 
Consequently, the operators involving \( B_{\mu\nu} \) and \( W_{\mu\nu}^A \) translate into effective couplings between the ALP and the physical \( \gamma \) and \( Z \) bosons:
\begin{eqnarray}
	\mathscr{L}_{\mathrm{eff}} \owns 
	- c_{\gamma\gamma} \frac{a}{f_a} F_{\mu\nu} \tilde{F}^{\mu\nu}
	- c_{\gamma Z} \frac{a}{f_a} F_{\mu\nu} \tilde{Z}^{\mu\nu}
	- c_{ZZ} \frac{a}{f_a} Z_{\mu\nu} \tilde{Z}^{\mu\nu},
\label{lagFZ}
\end{eqnarray}
where \( F_{\mu\nu} \) and \( Z_{\mu\nu} \) denote the electromagnetic and \( Z \) boson
field strength tensors, respectively. The Wilson coefficients \( c_{\gamma\gamma} \), \( c_{\gamma Z} \), and \( c_{ZZ} \) are given by:
\begin{eqnarray}
	c_{\gamma\gamma} = c_\theta^2 c_{BB} + s_\theta^2 c_{WW}, \quad 
	c_{\gamma Z} = s_{2\theta} (c_{WW}-c_{BB}), \quad 
	c_{ZZ} = s_\theta^2 c_{BB} + c_\theta^2 c_{WW},
\label{wilsoncoeff}
\end{eqnarray}
with \( s_\theta = \sin \theta_W \), \( c_\theta = \cos \theta_W \), and \( s_{2\theta} = 2 s_\theta c_\theta \), where \( \theta_W \) is the weak mixing angle. 
These relations show how the original couplings \( c_{BB} \) and \( c_{WW} \) control the ALP interactions with photons and \( Z \) bosons after EWSB. 

In general, when an ALP is produced, it may decay inside the detector (possibly reconstructed by its decay products) or leave the detector (manifesting itself as missing energy). The hadronic and leptonic decays of the ALP are not kinematically allowed here since the assumed ALP mass is well below the pion mass ($\approx 135$ MeV) and the on-shell electron pair production threshold ($2m_e \approx 1.022$ MeV). The decay into a pair of photons is, therefore, the dominant decay mode for the assumed ALP. At the leading order, the decay rate of an ALP into a pair of photons is given by
\begin{equation} 
\Gamma_{a \rightarrow \gamma\gamma} = \frac{m_a^3}{4\pi} \Bigl(\frac{c_{\gamma\gamma}}{f_a}\Bigr)^2.
\label{widtha2gaga}
\end{equation} 
As seen, the only Wilson coefficient contributing to this decay rate is ${c_{\gamma\gamma}}$. Other Wilson coefficients in the effective Lagrangian may also contribute to the decay of an ALP into a pair of photons if one takes into account beyond the tree-level effects. At the one-loop level, all the Wilson coefficients in the effective Lagrangian except for $c_{GG}$ contribute to the ALP decay to photons via fermion loops and electroweak loops. The ALP-gluon coupling starts to contribute to the ALP decay to photons at two-loop level. To take into account loop-induced effects, $c_{\gamma\gamma}$ in Eq. \ref{widtha2gaga} can be replaced with the effective Wilson coefficient $c_{\gamma\gamma}^\mathrm{eff}$ which includes higher-order effects~\cite{Bauer:2017ris}. The effective coefficient $c_{\gamma\gamma}^\mathrm{eff}$ depends on the ALP mass. However, this dependence is so mild that the decay rate scales as $m_a^3$ to an excellent approximation. The probability that an ALP decays inside the detector, $P_a^{\mathrm{\,det}}$, is given by
\begin{equation} 
P_a^{\mathrm{\,det}}=1-e^{-L_{\mathrm{det}}/L_a^\perp(\theta)}, 
\label{adecaylengthprob}
\end{equation} 
where $L_{\mathrm{det}}$ represents the transverse distance of the calorimeter from the collision point, $\theta$ is the angle between the ALP's velocity and the beam axis and $L_a^\perp$ denotes the decay length of the ALP perpendicular to the beam axis given by
\begin{equation} 
L_a^\perp(\theta) = \gamma\beta\tau\sin\theta = \frac{\sqrt{\gamma^2-1}}{\Gamma_a}\sin\theta \equiv L_a\sin\theta, 
\label{}
\end{equation} 
where $\gamma$ is the Lorentz boost factor, $\tau$ is the ALP proper lifetime, $\Gamma_a$ is the ALP width, and $\beta$ is the ALP speed. For small decay widths, the ALP escapes the detector before its decay leading to a large missing energy. The Lorentz boost factor, which depends on the experimental setup, also affects the detectability of ALPs. An ALP with mass 1 MeV can experience the visible decays $a\rightarrow\gamma\gamma$ and $a\rightarrow\gamma\nu\bar{\nu}$. However, the decay mode $a\rightarrow\gamma\nu\bar{\nu}$, mediated by the $a$-$Z$-$\gamma$ interaction, is negligible compared with the $a\rightarrow\gamma\gamma$ mode. This is because the limit on $\Gamma(a\rightarrow\gamma\nu\bar{\nu})$, deduced from the experimental limit on the Wilson coefficient $c_{\gamma Z}$~\cite{Brivio:2017ije}, is many orders of magnitude smaller than that of $\Gamma(a\rightarrow\gamma\gamma)$~\cite{Bauer:2017ris,Bauer:2017nlg}. The total decay width of such an ALP can be therefore approximated by the decay width of the ALP into a di-photon. Consequently, the ALP decay length in the laboratory frame can be estimated as
\begin{equation} 
L_a \approx \frac{|\vec{p}_a|}{m_a}\frac{1}{\Gamma_{a \rightarrow \gamma\gamma}} = \frac{4\pi}{m_a^4}\Bigl(\frac{ f_a}{c_{\gamma\gamma}}\Bigr)^2 |\vec{p}_a|,
\label{estimate1}
\end{equation}  
where $\vec{p}_a$ is the ALP three-momentum. Based on the current experimental limits, the parameter space region $c_{\gamma\gamma}/f_a>1\times10^{-9}\,\, \mathrm{TeV^{-1}}$ has been experimentally excluded for an ALP of mass 1 MeV \cite{Bauer:2017ris,Bauer:2017nlg}. Using Eq.~\ref{estimate1} and the upper limit on $c_{\gamma\gamma}/f_a$, one can obtain the lower limit
\begin{equation} 
L_a \gtrsim \left(\frac{2.3\,|\vec{p}_a|}{\mathrm{GeV}}\right) 10^{21} \,\mathrm{m}.
\label{estimate2} 
\end{equation}
on the ALP decay length. The ALP momentum $|\vec{p}_a|$ depends on many factors, e.g. the collider center-of-mass energy, triggers used in the analysis, event selection criteria imposed to enrich the signal, etc. Based on the Monte Carlo simulations performed in this study (described in section \ref{sec:eventgeneration}), the magnitude of the ALP momentum is of $\mathcal{O}(100)$ GeV for the assumed experimental setup. As a result, one can find the lower limit $L_a \gtrsim 2.3\times10^{23}\,\mathrm{m}$ on the ALP decay length using Eq.~\ref{estimate2}. Using the obtained lower limit in Eq. \ref{adecaylengthprob}, it can be seen that ALPs assumed in this work are likely to escape the detector and manifest themselves as missing energy. For ALPs  lighter than 1 MeV, the decay length becomes larger leading to more tendency to be invisible at the detector. It can be concluded that a large fraction of ALPs produced at the collider are invisible leading to a large missing energy signature. A very small fraction of ALPs decay inside the detector. The effect of the decaying ALPs is taken into account in this study. It is worth noting that for light enough ALPs, the lifetime can become larger than the age of the universe thanks to the third power of the ALP mass in Eq. \ref{widtha2gaga}. Such ALPs can be viable candidates for DM.

%***********************************************************
\section{ALP production in association with electroweak gauge bosons} 
\label{sec:signalandbkg}
\begin{figure}[t]
\centering
    \begin{subfigure}[b]{0.38\textwidth} 
    \centering
    \includegraphics[width=\textwidth]{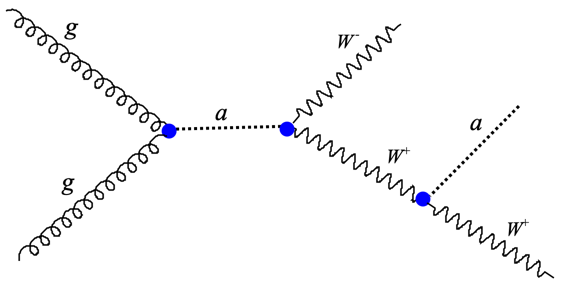}
    \end{subfigure} 
    \begin{subfigure}[b]{0.3\textwidth} 
    \centering
    \includegraphics[width=\textwidth]{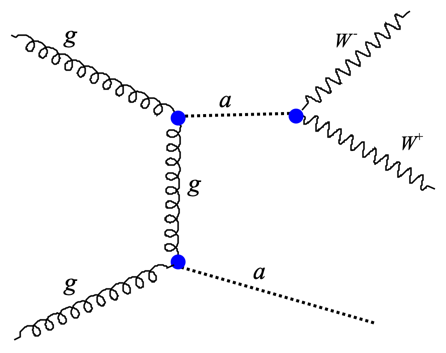}
    \end{subfigure} 
    \begin{subfigure}[b]{0.3\textwidth} 
    \centering
    \includegraphics[width=\textwidth]{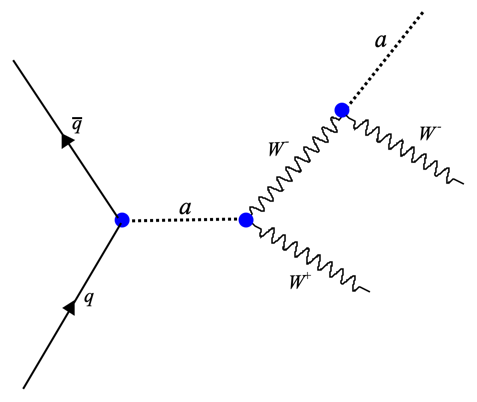}
    \end{subfigure} 
\par\bigskip
    \begin{subfigure}[b]{0.38\textwidth} 
    \centering
    \includegraphics[width=\textwidth]{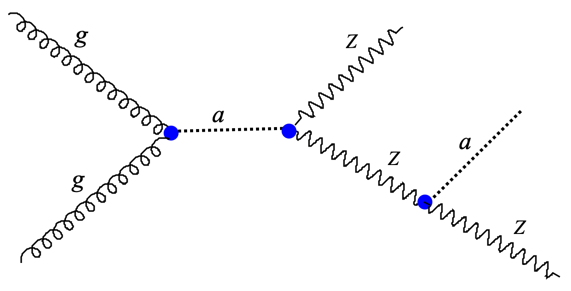}
    \end{subfigure}
    \begin{subfigure}[b]{0.3\textwidth} 
    \centering
    \includegraphics[width=\textwidth]{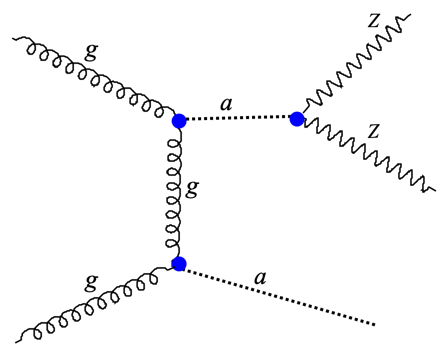}
    \end{subfigure}
    \begin{subfigure}[b]{0.3\textwidth} 
    \centering
    \includegraphics[width=\textwidth]{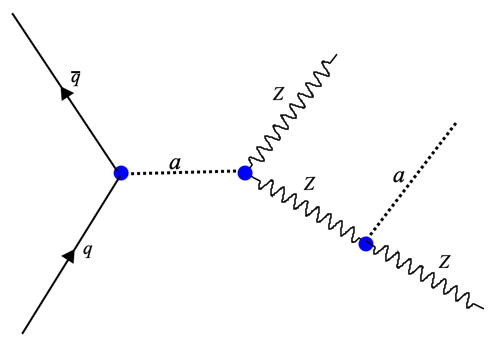}
    \end{subfigure}
  \caption{\small Representative Feynman diagrams contributing to the associated production of an ALP with two electroweak gauge bosons, $WW$ or $ZZ$, at the leading order at a proton-proton collider.}
\label{feynmanDiagrams}
\end{figure}

At a proton-proton collider, an ALP can be produced in association with two electroweak gauge bosons, $pp\rightarrow a VV$, where $VV$ can be $ZZ$ or $WW$. Fig. \ref{feynmanDiagrams} shows the representative Feynman diagrams contributing to these production processes at the leading order.
BSM interactions are shown with the blue filled circles in the Feynman diagrams. These processes are governed by the effective Lagrangian, Eq. \ref{lagfinal}. 
The interactions $a\mhyphen G\mhyphen G$, $a\mhyphen W\mhyphen W$, $a\mhyphen Z\mhyphen Z$ and $a\mhyphen q\mhyphen q$ contribute to these processes, 
and thus these processes are sensitive to the Wilson coefficients $c_{GG}$, $c_{WW}$, $c_{BB}$ and $c_{a\Phi}$ ($c_{ZZ}$ is a function of $c_{WW}$ and $c_{BB}$ 
as seen in Eq. \ref{wilsoncoeff}). Assuming the HL-LHC operating at the center-of-mass energy of 14 TeV, we study these processes and set upper bounds on the involved 
Wilson coefficients for an ALP of mass 1 MeV.

The processes considered in this analysis provide an opportunity to simultaneously 
probe both the ALP--gluon and ALP--electroweak couplings through a distinctive 
BSM topology involving three effective vertices. For example, the $ZZa$ signal process, 
in which the ALP interacts with gluons and electroweak gauge bosons, 
proceeds predominantly via gluon fusion, producing an  ALP in association with two $Z$ bosons. 
This results in characteristic kinematic features such as large invariant mass $m_{ZZ}$ and central rapidities. 
For the $ZZa$ final state, the squared matrix element for the dominant $t$-channel contribution 
scales as $|\mathcal{M}|^2 \propto \frac{c_{GG}^4 c_{ZZ}^2}{f_a^6} \times \hat{s}^2$, 
where $\hat{s}$ is the partonic center-of-mass energy. 
The corresponding cross section is enhanced by both the energy dependence 
of the squared matrix element and the steep rise of gluon PDFs at low Bjorken-$x$.

In contrast, there exists another process, $q\bar{q} \rightarrow ZZ a$, which results in a 
similar final state but proceeds through ALP radiation from a final-state $Z$ boson via 
an effective $aZ\tilde{Z}$ vertex, involving only a single non-zero ALP coupling. 
Its squared matrix element scales as $|\mathcal{M}|^2 \propto \frac{g^4 c_{ZZ}^2}{f_a^2} \times \hat{s}$. 
However, this contribution is significantly suppressed in the high-$m_{ZZ}$ region due to phase-space 
limitations and the rapidly falling quark PDFs at large momentum fractions. 
Moreover, the $Z$ bosons and ALP produced in this process tend to be softer 
and more forward than those in the BSM signal, resulting in much lower acceptance under 
our selection criteria, which favor large $m_{ZZ}$ and central leptons. 
Beyond these kinematic suppressions, the numerical impact of this process is negligible. 
For instance, for benchmark values $c_{GG} = c_{WW} = 0.5\ \text{TeV}^{-1}$, the cross section 
for the BSM-induced $ZZa$ process is approximately $2.27\ \text{pb}$, while the cross section 
for $q\bar{q} \rightarrow ZZ a$ is only $0.0122\ \text{pb}$ (with $c_{WW} = 0.5\ \text{TeV}^{-1}$), corresponding to less 
than $1\%$ of the signal rate. Thus, even if included, the $q\bar{q} \rightarrow ZZ a$ 
process would have no meaningful impact on our signal region.
Furthermore, the $q\bar{q} \rightarrow ZZ a$ topology probes only a single coupling (e.g., $c_{ZZ}$) and 
contributes at $\mathcal{O}(c_i^2)$, dominating in the low-coupling regime. 
Including such terms would bias the interpretation of the exclusion limits by reducing 
sensitivity to the interplay between multiple couplings and leading to 
exclusion contours dominated by a single direction in parameter space. 
This would undermine the central objective of simultaneously constraining multiple ALP interactions in a model-independent framework.
By restricting our analysis to signal diagrams that involve three effective ALP vertices---and which do not interfere with SM 
amplitudes at leading order---we ensure a clean, interpretable extraction of two-dimensional 
exclusion limits in the $(c_{GG}, c_{WW})$ or $(c_{GG}, c_{ZZ})$ planes.

\begin{figure}[ht]
\centering
    \centering
    \includegraphics[width=0.45\textwidth]{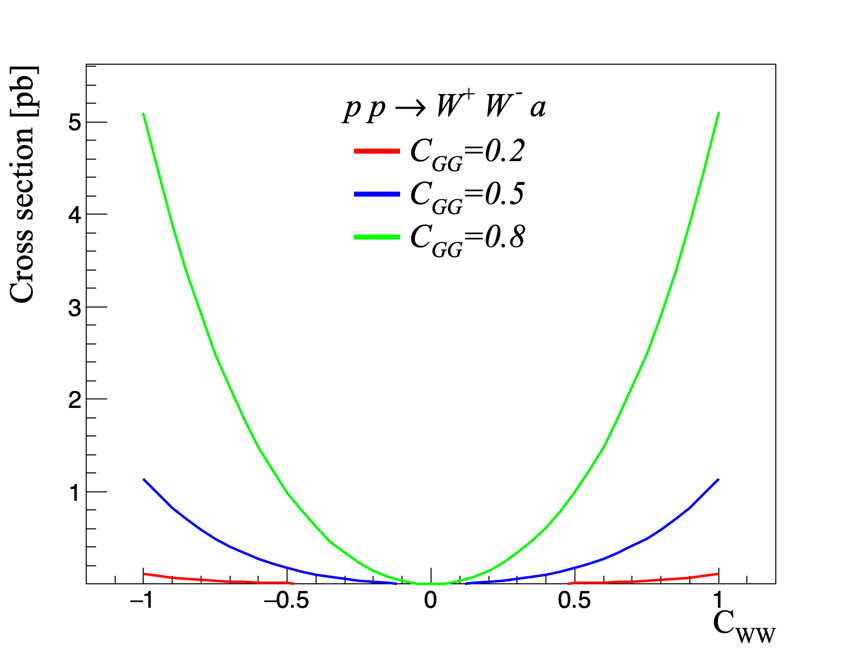}
     \includegraphics[width=0.45\textwidth]{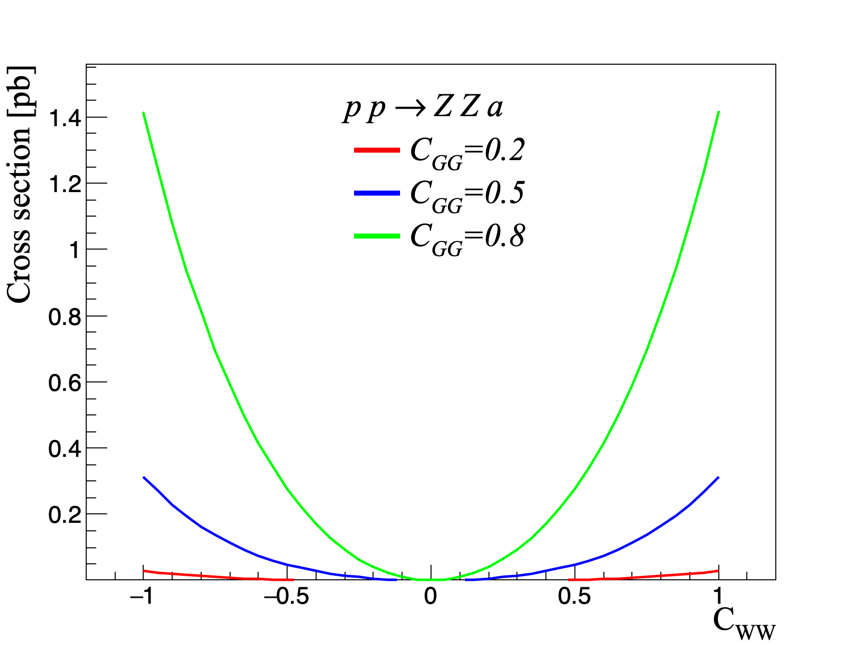}\\
   \caption{\small Leading-order cross sections of the processes $pp\rightarrow WWa$ (left) and $pp\rightarrow ZZa$ (right) 
   as a function of $c_{WW}$ assuming $\sqrt{s}=14$ TeV, $m_a=1$ MeV and two non-zero Wilson coefficients at a time.}
\label{signalxsec}
\end{figure} 

As shown in Fig.\ref{signalxsec}, the $WWa$ and $ZZa$ production processes exhibit greater sensitivity to the 
coupling $c_{GG}$ compared to $c_{WW}$. Notably, both processes demonstrate minimal sensitivity to the 
coupling $c_{a\Phi}$ which arise from sub-processes of $q\bar{q}$ in the initial state. 
For instance, in $WWa$ production, when $c_{WW}$ and $c_{a\Phi}$ are fixed at 1.0, the corresponding cross section
 is of the order of $10^{-6}$. 
 The main reason for such a small cross section is that the ALP coupling to quarks (fermions) is proportional to 
 the Yukawa coupling constants (see Eq. \ref{lagfinal}). 
 The smallness of the Yukawa coupling constants corresponding to the up and down quarks significantly 
 suppresses the cross section of sub-processes involving the ALP coupling to quarks. 
 Hence, sub-processes involving the ALP coupling to gluons make the dominant contribution to 
 the $WWa$ and $ZZa$ production with respect to the sub-processes with $q\bar{q}$ in the initial state
 hence the presence of  $c_{a\Phi}$ unlikely  produces any discernible experimental effect, justifying its omission from further detailed exploration in this study.

\begin{figure}[ht]
\centering
    \centering
    \includegraphics[width=0.55\textwidth]{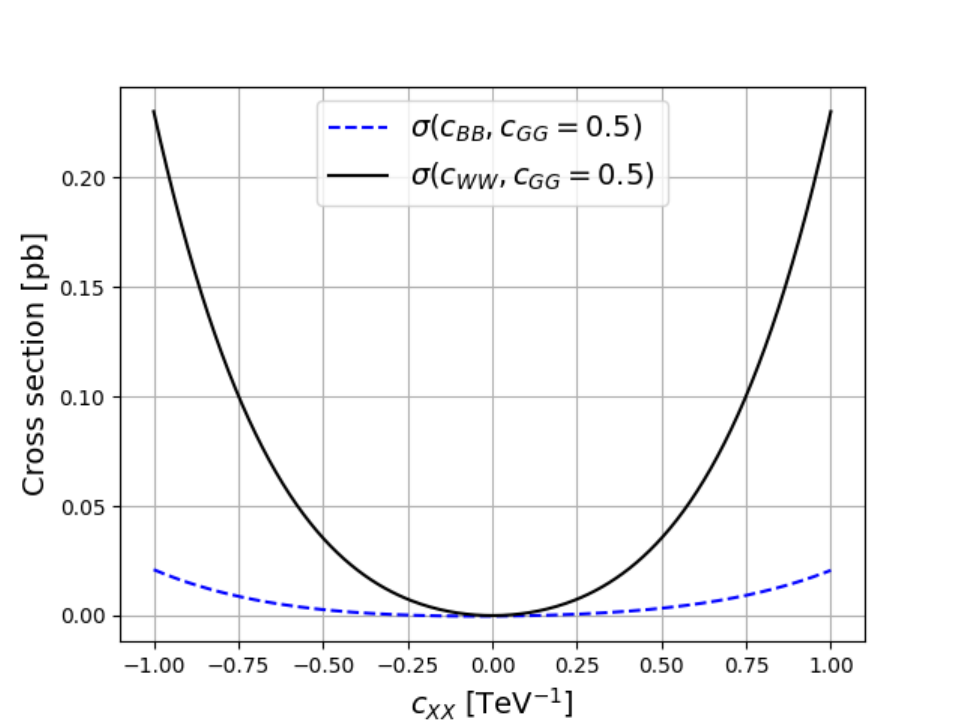}
   \caption{\small Leading-order production cross section of the  $pp\rightarrow ZZa$ 
   as a function of $c_{WW}$ and $c_{BB}$ assuming $c_{GG} = 0.5$ TeV$^{-1}$ and $m_a=1$ MeV.}
\label{cscBB}
\end{figure}

The production cross section for the process $pp \rightarrow ZZa$ exhibits a dependence on the Wilson coefficient $c_{BB}$, 
as detailed in Eqs. \ref{lagFZ} and \ref{wilsoncoeff}. Figure \ref{cscBB} illustrates the variation of the cross section 
$\sigma(pp \rightarrow ZZa)$ as a function of both $c_{BB}$ and $c_{WW}$, under the assumption that the gluonic Wilson coefficient is fixed at $c_{GG} = 0.5~\text{TeV}^{-1}$.
A key observation is that the sensitivity to $c_{BB}$ is notably weaker compared to $c_{WW}$. 
This can be attributed to the specific coupling structure inherent to the interaction terms, particularly 
the presence of a suppression factor proportional to $\sin^2 \theta_W$ (denoted $s_{\theta}^2$) in the $c_{BB}$ 
contribution to $c_{ZZ}$. In contrast, the $c_{WW}$ contribution involves a factor of $\cos^2 \theta_W$
 (denoted $c_{\theta}^2$), which leads to a comparatively stronger enhancement for $c_{WW}$ in processes involving neutral gauge bosons.
As a result, the variation in the production rate of $pp \rightarrow ZZa$ with respect to $c_{BB}$ remains minimal, and the overall 
cross section is both low and exhibits limited sensitivity to changes in $c_{BB}$. This reduced sensitivity suggests that $c_{BB}$ does 
not significantly contribute to deviations in the cross section within the explored parameter space.
Therefore, while the impact of $c_{BB}$ on the cross section is presented and quantified, it remains
subdominant compared to $c_{WW}$ and no further detailed comparison is necessary for the 
purpose of setting the main exclusion limits.

A scan over the ALP mass shows that the production cross sections for both $WWa$ and $ZZa$ processes 
remain nearly independent of the ALP mass for values below a few GeV.
This is because the ALP mass in this range 
is small compared with the typical energy scale of the process. It is also seen that the kinematics of the $WWa$ and $ZZa$ 
events is roughly independent of the ALP mass for such light ALPs. As a result, the constraints obtained in this study 
(provided in section \ref{sec:constraint}) hold not only for an ALP of mass 1 MeV, but also for lighter ALPs to an excellent approximation.

The fully leptonic final state is considered for the $WWa$ signal in this study. For the $ZZa$ signal, the fully leptonic and semi-leptonic final states are studied. 
The dominant SM processes contributing to the background for the $WWa$ production in the fully leptonic final state and the $ZZa$ production in the semi-leptonic
 final state are assumed to be $t\bar{t}$, $tW$, $WW$, $WZ$, $ZZ$, $ttZ$, $WWZ$ and $ttH$. The dominant SM backgrounds for the $ZZa$ production in the fully 
 leptonic final state are assumed to be $ZZ$, $ttZ$, $WWZ$, $ttH$, $WWWW$ and $tttt$. In these SM processes, neutrinos coming from jets or leptonic decays 
 of $W$ and $Z$ bosons cause missing energies that mimic the signal missing energy signature.

\section{Analysis} 
\label{sec:analysis}
In this section, we discuss the details of the analysis. The analysis begins by the generation of the signal and background events. In the simulated events, 
objects such as jets and isolated leptons are identified. Then, events are selected and analyzed to separate the signal from background. 
A condition is also imposed to insure the validity of the effective Lagrangian.

%***********************************************************
\subsection{Event generation} 
\label{sec:eventgeneration}
The effective Lagrangian, Eq. \ref{lagfinal}, has been implemented into \textsc{FeynRules} \cite{Alloul:2013bka} and the generated Universal FeynRules Output (UFO) \cite{Degrande:2011ua} model has been passed to \textsc{MadGraph5\_aMC@NLO}~\cite{Alwall:2011uj} to compute the cross section and generate hard events for the signal and background processes. NNPDF23 \cite{Ball:2012cx} is used as the proton PDF and the center-of-mass energy of the proton beams is set to 14 TeV. \textsc{Pythia 8.2.43} is used to perform parton showering, hadronization and decays of unstable particles. 
Showered events are then passed to \textsc{Delphes} 3.4.2~\cite{deFavereau:2013fsa} for fast detector simulation, using the official CMS detector card \texttt{delphes\_card\_CMS.tcl}\footnote{\url{https://github.com/delphes/delphes/blob/master/cards/delphes_card_CMS.tcl}} provided in the \textsc{Delphes} distribution.
While the present study relies on fast detector simulation using \textsc{Delphes} with the CMS detector card, 
a future extension of this work foresees the use of a full \textsc{GEANT4}-based simulation \cite{geant} to 
more accurately model jet reconstruction and low-energy particle effects, thereby validating the robustness of the present results. 

The signal samples were generated at leading order using the ALP linear 
UFO model\footnote{\url{https://feynrules.irmp.ucl.ac.be/attachment/wiki/ALPsEFT/ALP_linear_UFO.tar.gz}}, 
which implements the effective interactions of the ALP with SM gauge bosons in the linear EFT framework. 
All SM background processes were also generated at leading order.
The generation of $WWa$ and $ZZa$ signal samples has been performed assuming two non-zero Wilson coefficients, $c_{WW}$ and $c_{GG}$, at a time. 
The $WWa$ and $ZZa$ samples are both used to derive exclusion limits in the $(c_{WW}, c_{GG})$ plane.
 The ALP mass, $m_a$, is set to 1 MeV for all samples. As discussed in section \ref{sec:signalandbkg}, a very small fraction of ALPs produced in the collider decay inside the detector. The effect of the decaying ALPs is taken into account in an event-by-event manner with the use of the decay probability, $P_a^{\mathrm{\,det}}$, defined in Eq.~\ref{adecaylengthprob}.

%***********************************************************
\subsection{Validity of the ALP effective Lagrangian} 
\label{subsec:validity}
To ensure the validity of the effective Lagrangian, the mass scale of new physics, $f_a$, should be significantly larger than the typical energy scale, $\sqrt{\hat{s}}$, of the process under study. The strict requirement for the validity of the effective description is, therefore, to satisfy the condition $\sqrt{\hat{s}}<f_a$. However, $\sqrt{\hat{s}}$ cannot be experimentally measured due to the presence of invisible ALPs in the final state. One may use the correlation between $\sqrt{\hat{s}}$ and the missing transverse energy, obtained using the Monte Carlo simulated events, to naively ensure validity of the EFT by requiring $2\slashed{E}_T^{max}<f_a$, where $\slashed{E}_T^{max}$ is the highest missing transverse energy data bin in the analysis. In this work, the strict EFT validity condition $\sqrt{\hat{s}}<f_a$ is imposed by discarding events that do not satisfy the condition $\sqrt{\hat{s}}<2\slashed{E}_T^{max}$.

%***********************************************************
\subsection{Object identification and event selection} 
\label{selection}
Generated $WWa$ and $ZZa$ samples are independently analyzed to constrain BSM couplings. 
Jet reconstruction is performed using the anti-$k_t$ algorithm~\cite{Cacciari:2008gp} implemented in \texttt{FastJet 3.3.2}~\cite{Cacciari:2011ma}, 
with a radius parameter $R = 0.4$ that defines the jet cone size in the rapidity-azimuthal angle ($y$-$\phi$) plane, such that $\Delta R = \sqrt{(\Delta y)^2 + (\Delta \phi)^2}$.
The transverse momentum and pseudorapidity of reconstructed jets are required to satisfy the conditions $p_T>30$ GeV and $\vert \eta \vert < 2.5$. 
Isolated electrons and muons are identified using the relative isolation variable, $I_{rel}$. This variable is defined as $I_{rel}=\sum p_T^{\,i}/p_T^{\,\mathrm{P}}$, where P is the candidate particle for which $I_{rel}$ is calculated, and $i$ is the summation index running over all particles (excluding the particle P) within a cone of radius $R = 0.5$ centered on the candidate particle, defined in the pseudorapidity-azimuthal angle plane as $\Delta R = \sqrt{(\Delta \eta)^2 + (\Delta \phi)^2}$.
All particles with a transverse momentum greater than 0.5 GeV are taken into account in the summation. An electron (muon) is identified as an isolated lepton if $I_{rel}<0.12$ ($0.25$). Isolated electrons and muons should satisfy the conditions $p_T>20$ GeV and $\vert \eta \vert < 2.4$. 

To analyze the $WWa$ signal in the fully leptonic final state, events are required to have exactly two isolated oppositely-charged leptons (electron or muon) and no jet. The flavors of the leptons can be the same or different. For the $ZZa$ analysis, both the fully leptonic and semi-leptonic final states are studied. Two signal regions corresponding to these final states are defined as follows. SR1 is defined by the requirement that events should have exactly two pairs of isolated oppositely-charged same-flavor leptons (electron or muon) and no jet. This includes events with the final states $ee\mu\mu$, $eeee$ and $\mu\mu\mu\mu$. SR2 is defined by requiring exactly one pair of isolated oppositely-charged same-flavor leptons ($ee$ or $\mu\mu$) and at least two jets. The missing transverse energy, $\slashed{E}_T$, which shows the energy imbalance in the transverse plane is required to be larger than 30 GeV for all events for the $ZZa$ SR2 and $aWW$ analyses.  In the analysis, jets from all parton flavors are considered, with no flavor-based selection, 
and jets found within a cone of $\Delta R < 0.4$ around any selected lepton are removed to avoid overlap. 
In  SR1, lepton pairs are required to have invariant masses consistent with the $Z$ boson mass window. 
Fake leptons and jet misidentification effects are not included in the fast detector simulation via Delphes and are therefore beyond the scope of this study.

Applying the above-mentioned cuts, event selection efficiencies presented in Tab.~\ref{eff} are obtained for the signal and background samples. As seen, the SM production of $WWWW$, $ZZ$ and $tttt$ have the highest selection efficiencies in the SR1 signal region of the $ZZa$ analysis. For the SR2 signal region, the processes $ttZ$, $WWZ$ and $t\bar{t}$ have the highest efficiencies. Finally, the processes $tW$, $t\bar{t}$ and $WW$ are the most important backgrounds in the $aWW$ analysis.
\begin {table}[t] 
%1
\begin{subtable}{\textwidth}
\centering
         \begin{tabular}{cccccccc} 
 & $ZZa\,(\mathrm{SR1})$ & $ZZ$ & $ttZ$ & $WWZ$ & $ttH$ & $WWWW$ & $tttt$  
  \parbox{0pt}{\rule{0pt}{1ex+\baselineskip}}\\ \Xhline{1\arrayrulewidth} 
   & 0.071  & 0.0024 & 0.00043 & 0.00047 & 1.5e-6 & 0.0035 & 0.0013 \parbox{0pt}{\rule{0pt}{1ex+\baselineskip}}\\
        \end{tabular} 
\label{effdi}
%\caption {}
%\label{eff}
\medskip
\end{subtable}
%2
\begin{subtable}{\textwidth}
\centering
         \begin{tabular}{cccccccccc} 
 & $ZZa\,(\mathrm{SR2})$ & $t\bar{t}$ & $tW$ & $WW$ & $WZ$ & $ZZ$ & $ttZ$ & $WWZ$ & $ttH$  
  \parbox{0pt}{\rule{0pt}{1ex+\baselineskip}}\\ \Xhline{1\arrayrulewidth} 
   & 0.396  & 0.034 & 0.024 & 0.0022 & 0.013 & 0.015 & 0.082 & 0.043 & 0.0028 \parbox{0pt}{\rule{0pt}{1ex+\baselineskip}}\\
        \end{tabular} 
\label{effdi}
%\caption {}
%\label{eff}
\medskip
\end{subtable}
%3
\begin{subtable}{\textwidth}
\centering
         \begin{tabular}{cccccccccc} 
 & $WWa$ & $t\bar{t}$ & $tW$ & $WW$ & $WZ$ & $ZZ$ & $ttZ$ & $WWZ$ & $ttH$  
  \parbox{0pt}{\rule{0pt}{1ex+\baselineskip}}\\ \Xhline{1\arrayrulewidth} 
   & 0.478  & 0.199 & 0.235 & 0.128 & 0.00091 & 0.0017 & 0.096 & 0.113 & 0.013 \parbox{0pt}{\rule{0pt}{1ex+\baselineskip}}\\
        \end{tabular} 
\label{effdi}
%\caption {}
%\label{eff}
\end{subtable}
\caption{\small Event selection efficiencies after applying all selection cuts obtained for the signal ($ZZa$, $WWa$) and different background processes. Two efficiencies are presented for each signal process. For the signal processes the couplings values are $c_{WW}=0.1, \, c_{GG}=0.1$.} 
\label{eff}
\end {table} 

\subsection{Signal-background discrimination} 
\label{discrimination}
\begin{figure*}[!ht]
  \centering  
    \begin{subfigure}[b]{0.49\textwidth} 
    \centering
    \includegraphics[width=\textwidth]{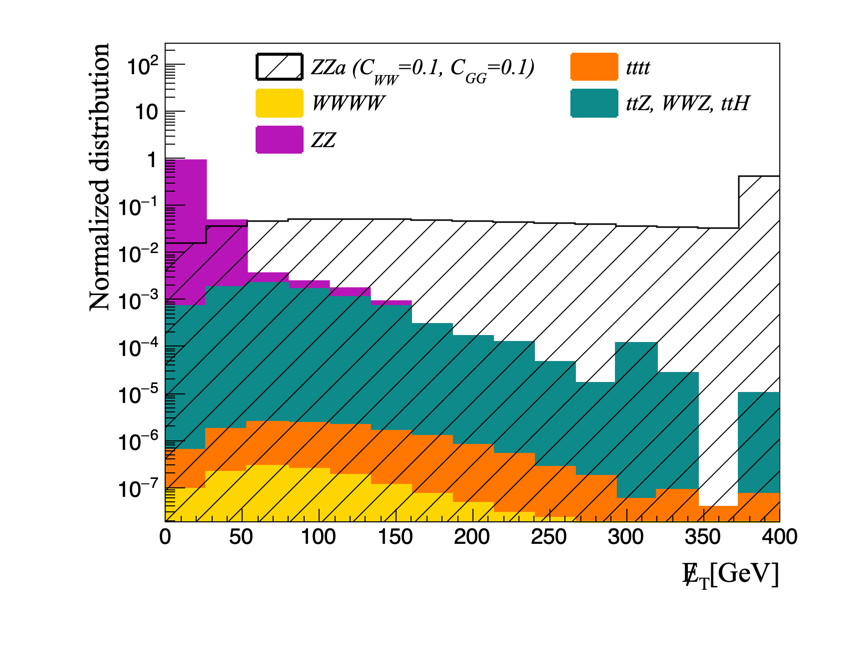}
    \caption{}
    \label{Met_azz}
    \end{subfigure} 
    \begin{subfigure}[b]{0.48\textwidth}
    \centering
    \includegraphics[width=\textwidth]{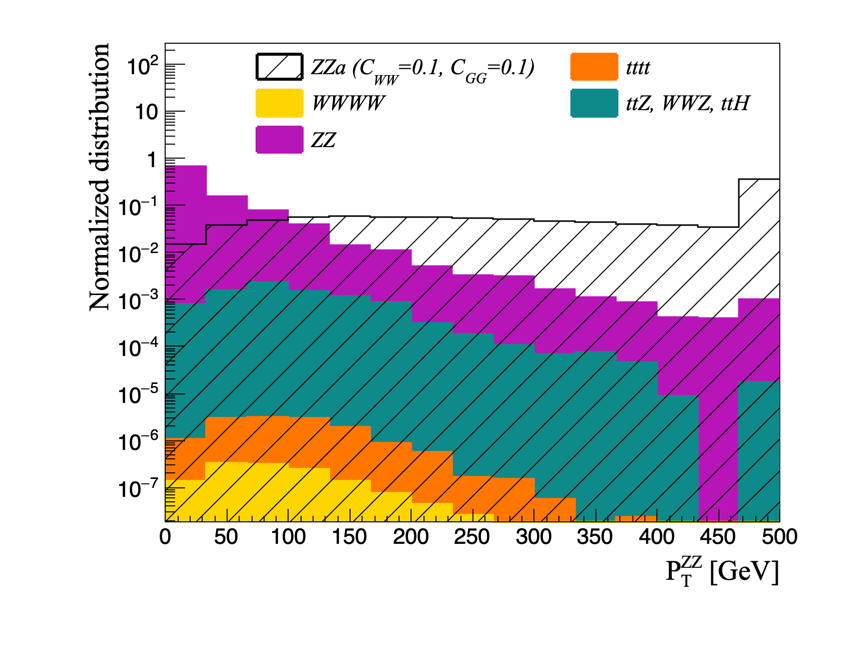}
    \caption{}
    \label{ptz1z2_azz}
    \end{subfigure} 
    \begin{subfigure}[b]{0.48\textwidth} 
    \centering
    \includegraphics[width=\textwidth]{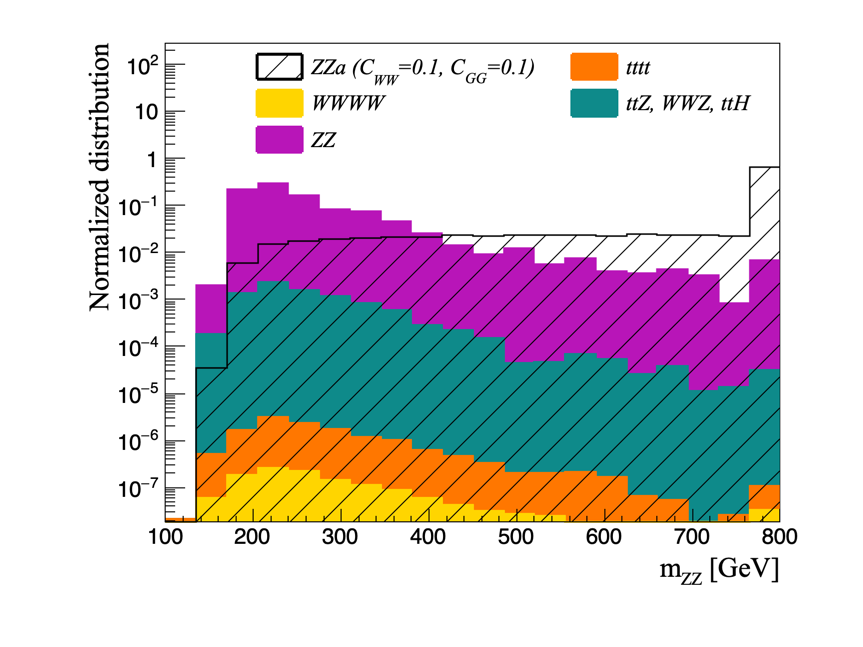}
    \caption{}
    \label{mzz_azz}
    \end{subfigure}
    \begin{subfigure}[b]{0.49\textwidth} 
    \centering
    \includegraphics[width=\textwidth]{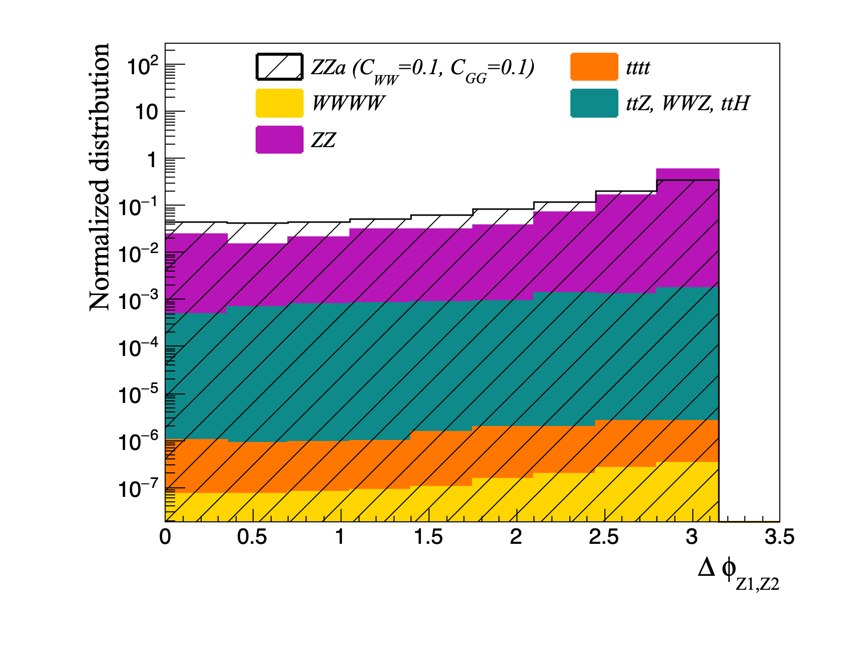}
    \caption{}
    \label{Deltaphiz1z2_azz}
    \end{subfigure} 
\caption{\small Distributions of the discriminating variables for the signal and background processes obtained for the SR1 signal region in the $ZZa$ analysis. The signal and the total background distributions are normalized to unity.}
\label{azz_variables}
\end{figure*}

Signal events with non-zero $c_{WW}$ and $c_{GG}$ are analyzed independently for the $ZZa$ (SR1 and SR2) and $WWa$ 
processes resulting in exclusion limits in the $c_{WW}$-$c_{GG}$  plane, respectively. 
Different variables are defined and used to discriminate between the selected events in the signal and background samples. 
To obtain the best signal-background discrimination, a multivariate technique using the TMVA (Toolkit for Multivariate Data Analysis) package \cite{Hocker:2007ht,Speckmayer:2010zz,Therhaag:2010zz} is deployed. The discrimination power of all the multivariate classification algorithms 
available in the TMVA package are compared using the receiver operating characteristic (ROC) curve to find the algorithm with the highest discrimination power. 
The Boosted Decision Trees (BDT) algorithm \cite{Xia:2018cfz} was found to be the most powerful algorithm for both the $ZZa$ and $WWa$ analyses and 
is thus used in this study. The distributions obtained for the discriminating variables are passed to the BDT algorithm, and the BDT performs the training process 
considering all background processes according to their respective weights. The TMVA overtraining check is performed to ensure overtraining does not occur. 
Kolmogorov-Smirnov (K-S) test is also performed to ensure the consistency of the BDT responses obtained for the training and test samples. 
The discriminating variables defined to be used for the SR1 signal region in the $ZZa$ analysis are:
\begin{itemize}    
    \item Missing transverse energy, $\slashed{E}_T$.
    \item Invariant mass of the four reconstructed charged leptons, $m_{ZZ}$.
    \item Magnitude of the sum of transverse momentum vectors of the four reconstructed charged leptons, $p_T^{ZZ}$.
    \item Azimuthal separation between the two reconstructed $Z$ bosons, $\Delta\Phi_{Z_1Z_2}$. Each $Z$ boson is reconstructed using a pair of oppositely-charged same-flavor charged leptons. If two lepton pairs of the same flavor exist, i.e. $eeee$ and $\mu\mu\mu\mu$, the pair of oppositely-charged leptons with minimum $|m_{\ell\ell}-m_Z|$ (with $m_Z\approx91$ being the $Z$ boson mass) is used to reconstruct the first $Z$ boson ($Z1$), and the remaining pair is used to reconstruct the second $Z$ boson ($Z2$). 
\end{itemize} 
Distributions obtained from the selected events for the above variables are shown in Fig. \ref{azz_variables}.
According to the BDT output, $\slashed{E}_T$, $p_T^{ZZ}$ and $m_{ZZ}$ are the best variables in terms of the discrimination power for the $ZZa$ SR1 analysis. The discriminating variables listed below are used for the $WWa$ and $ZZa$ SR2 analyses:
 \begin{itemize}   
    \item Missing transverse energy, $\slashed{E}_T$.
    \item Invariant mass of the two reconstructed charged leptons, $m_{\ell\ell}$.
    \item Magnitude of the sum of transverse momentum vectors of the two reconstructed charged leptons, $p_T^{\ell\ell}$.
    \item Spatial separation, $\Delta R = \sqrt{\Delta\eta^2+\Delta\phi^2}$, with $\eta$ and $\phi$ being the pseudorapidity and azimuth angle, between the two reconstructed charged leptons, $\Delta R_{\ell_1\ell_2}$. 
    \item Number of reconstructed jets, $N_{jet}$.
\end{itemize}
\begin{figure*}[!ht]
  \centering  
    \begin{subfigure}[b]{0.49\textwidth} 
    \centering
    \includegraphics[width=\textwidth]{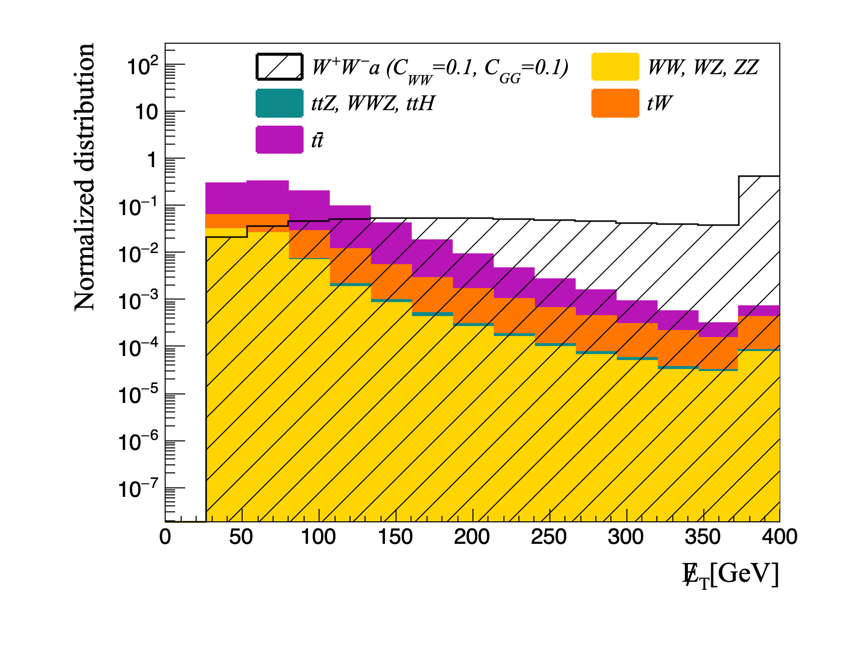}
    \caption{}
    \label{Met_aww}
    \end{subfigure} 
    \begin{subfigure}[b]{0.49\textwidth}
    \centering
    \includegraphics[width=\textwidth]{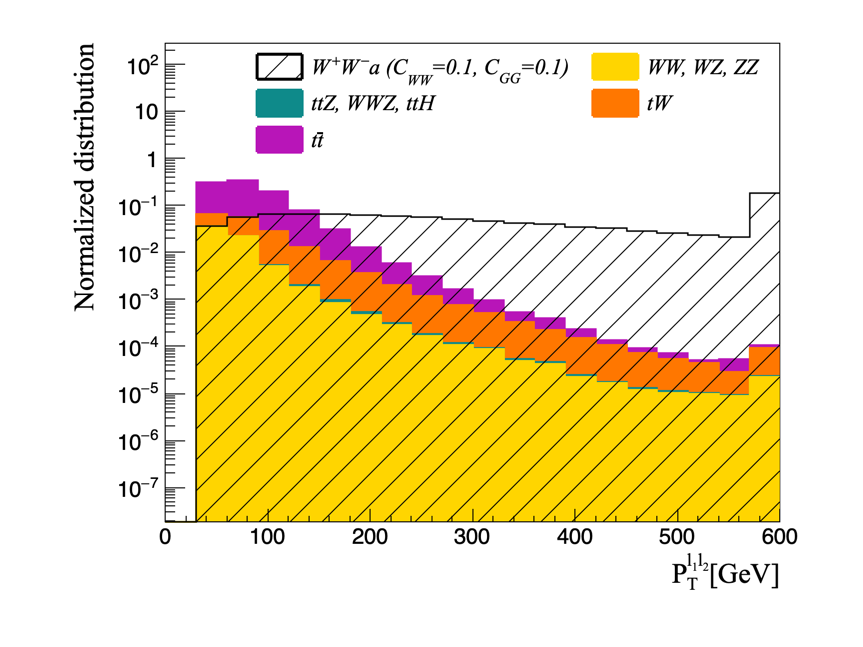}
    \caption{}
    \label{Ptl1l2_aww}
    \end{subfigure} 
    \begin{subfigure}[b]{0.49\textwidth} 
    \centering
    \includegraphics[width=\textwidth]{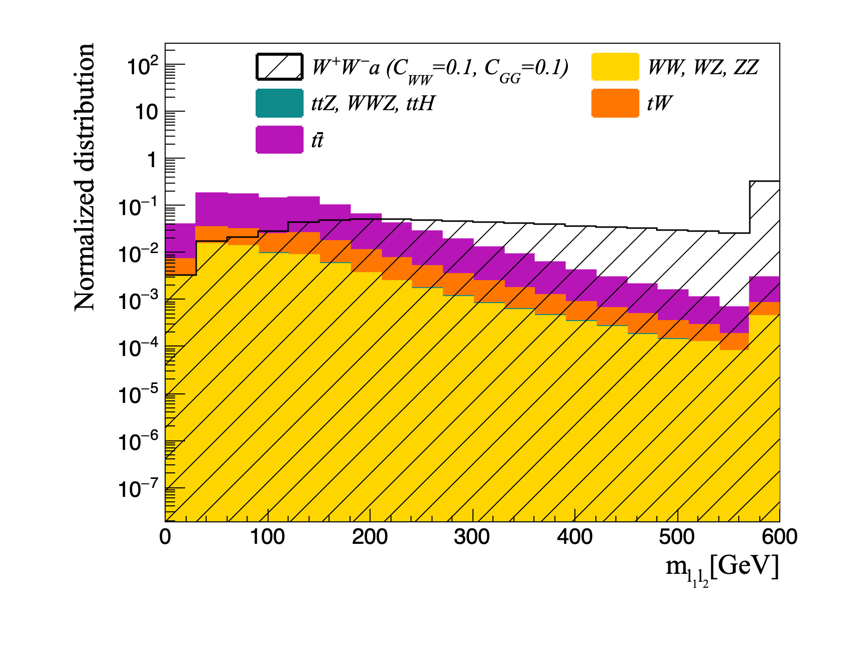}
    \caption{}
    \label{ml1l2_aww}
    \end{subfigure}
    \begin{subfigure}[b]{0.49\textwidth} 
    \centering
    \includegraphics[width=\textwidth]{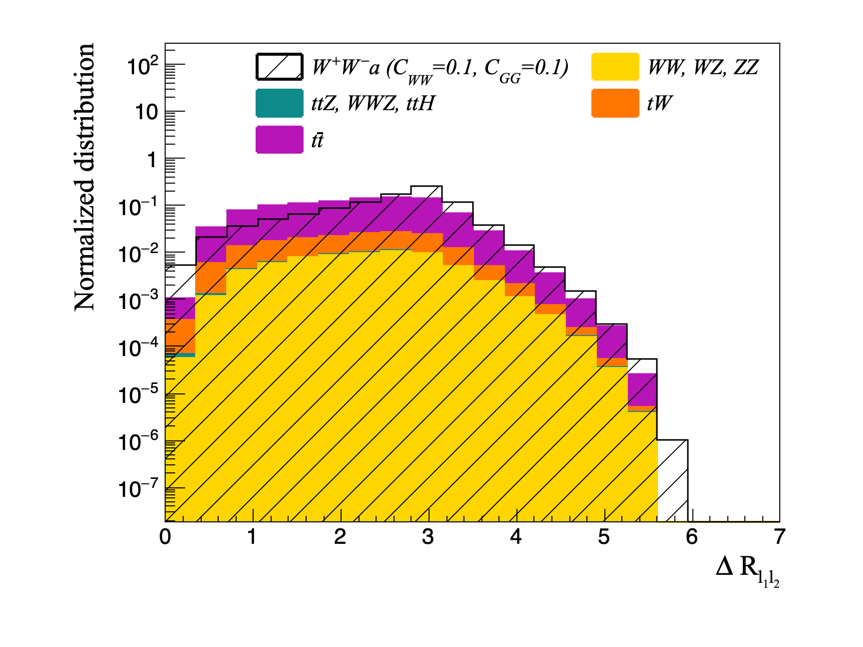}
    \caption{}
    \label{DR_l1l2_aww}
    \end{subfigure} 
    \begin{subfigure}[b]{0.49\textwidth} 
    \centering
    \includegraphics[width=\textwidth]{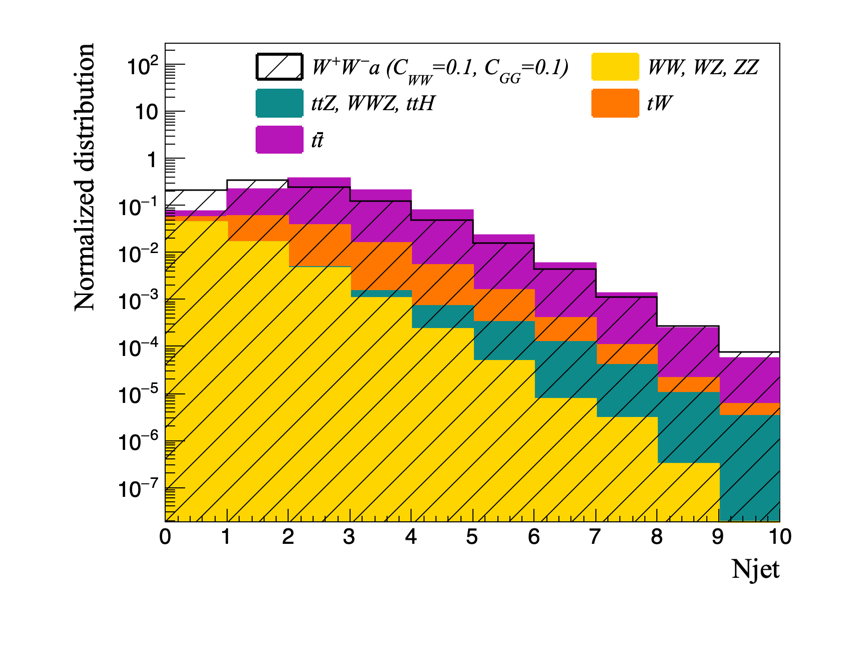}
    \caption{}
    \label{njet_aww}
    \end{subfigure}
\caption{\small Distributions of the discriminating variables for the signal and background processes obtained in the $WWa$ analysis. The signal and the total background distributions are normalized to unity.}
\label{aww_variables}
\end{figure*}
Distributions obtained for these variables in the $WWa$ analysis (for example) are shown in Fig. \ref{aww_variables}. 

As discussed in section \ref{sec:signalandbkg}, the dominant fraction of ALPs produced at the collider decay outside the detector. The resulting missing energy signature plays an important role in separating the signal from background in both the $WWa$ and $ZZa$ analyses. The only source of missing energy in the SM background is the missing neutrinos which result in missing energies typically lower than that of the signal (see Figs. \ref{Met_azz},\ref{Met_aww}). The magnitude of the sum of transverse momentum vectors of the reconstructed leptons, i.e. $p_T^{ZZ}$ and $p_T^{\ell\ell}$, are also powerful variables as the final state ALP in the $ZZa$ and $WWa$ processes prevents the produced $W$ and $Z$ bosons from being emitted in a back-to-back configuration (unlike the case of $ZZ$ and $WW$ production) resulting in relatively large momenta sum values. Furthermore, the dominance of the $s$-channel diagrams contribution to the signal cross section results in $W$ and $Z$ bosons mostly emitted near the transverse plane leading to high transverse momenta sums. As a result, the $ZZ$ and $WW$ production processes, which are respectively the dominant backgrounds for the $ZZa$ and $WWa$ signals, are significantly suppressed by the momenta sum variables (see Figs. \ref{ptz1z2_azz},\ref{Ptl1l2_aww}). 
As seen in Fig. \ref{njet_aww}, the number of jets, $N_{jet}$, is particularly useful for suppressing backgrounds with high jet multiplicities like $ttZ$, $WWZ$, $ttH$ and $tt$. According to the BDT output, $m_{\ell\ell}$, $\Delta R_{\ell_1\ell_2}$ and $\slashed{E}_T$ ($\slashed{E}_T$, $N_{jet}$ and $p_T^{\ell\ell}$) are the most powerful variables for the $ZZa$ SR2 ($WWa$) analysis to discriminate the signal from background. 

\begin{figure}[!ht]
\centering
    \begin{subfigure}[b]{0.49\textwidth} 
    \centering
    \includegraphics[width=\textwidth]{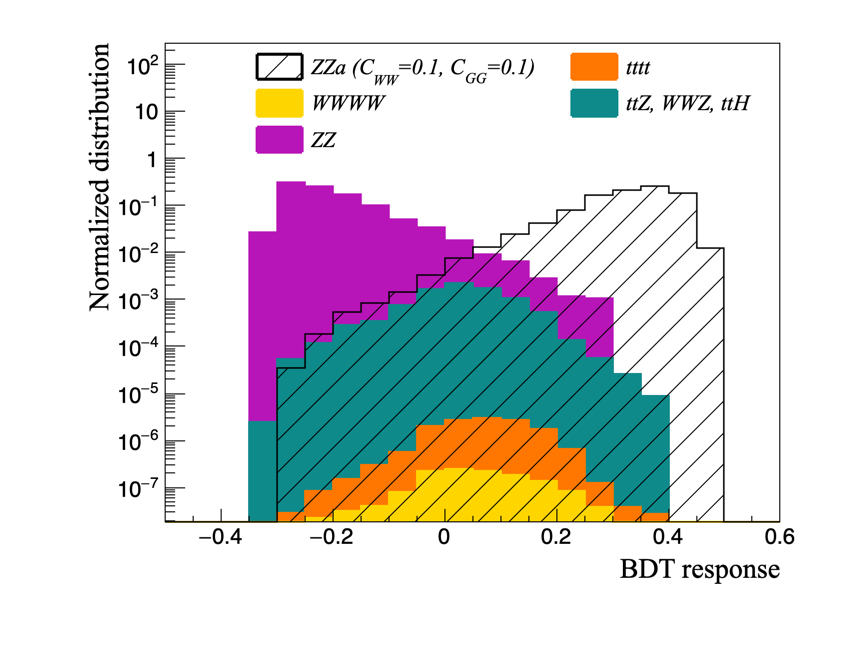}
    \caption{}
    \label{BDT1_ZZSR1}
    \end{subfigure} 
    \begin{subfigure}[b]{0.49\textwidth} 
    \centering
    \includegraphics[width=\textwidth]{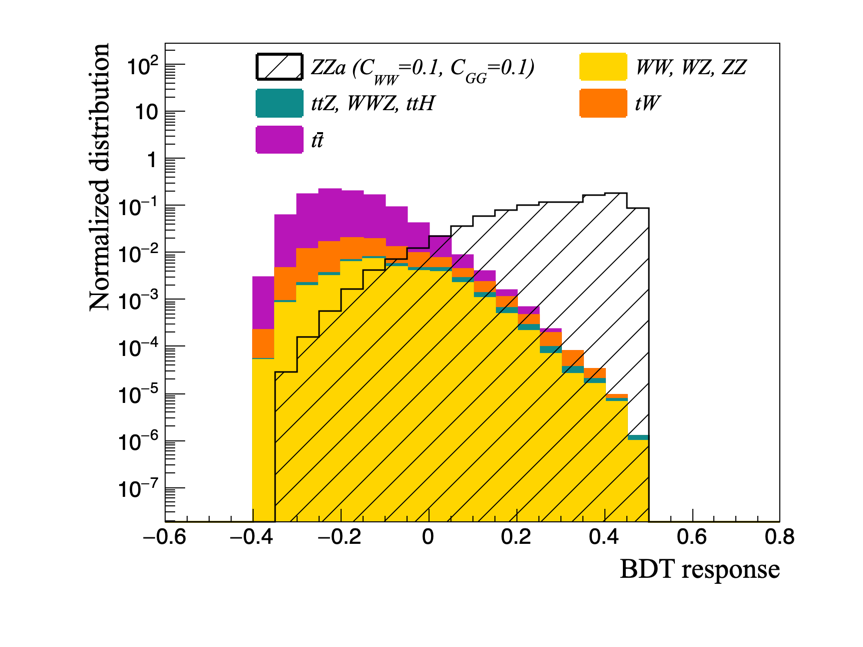}
    \caption{}
    \label{BDT1_ZZSR2}
    \end{subfigure}
    \begin{subfigure}[b]{0.49\textwidth} 
    \centering
    \includegraphics[width=\textwidth]{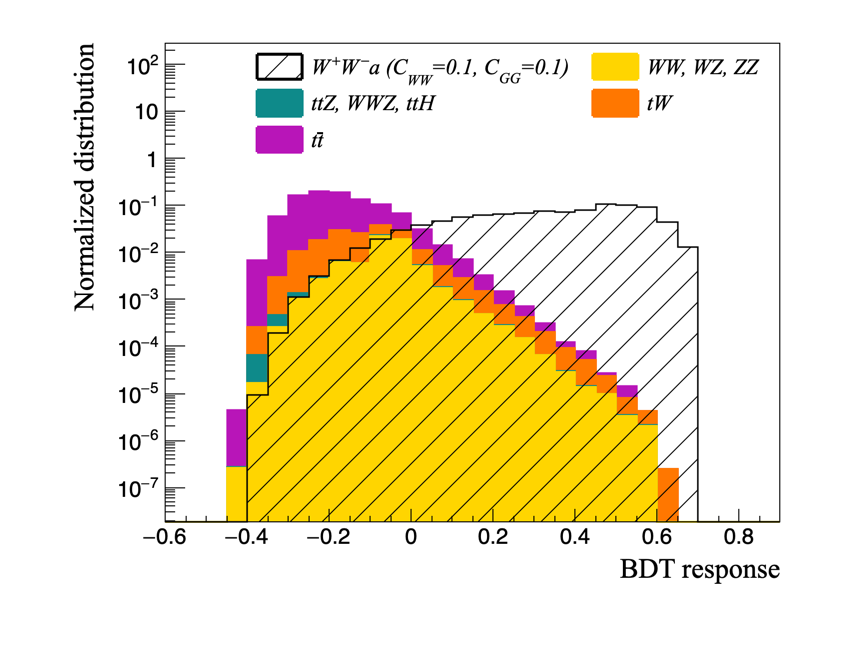}
    \caption{}
    \label{BDT1_WW}
    \end{subfigure}
  \caption{\small Distributions of the BDT response obtained for the signal and background processes 
  corresponding to the ALP mass of 1 MeV and $\sqrt{s}=14$ TeV for a) $ZZa$ SR1, b) $ZZa$ SR2 and c) 
  $WWa$ analyses. For each analysis, distributions are presented for the case of non-zero $c_{GG}$ and $c_{WW}$.}
\label{BDTs}
\end{figure}

Using the trained model, BDT responses for the signal and background events are obtained. Fig. \ref{BDTs} shows the obtained results for the $ZZa$ SR1, $ZZa$ 
SR2 and $WWa$ analyses. 
As seen, the BDT performs well in signal-background discrimination. 
A comparison shows that the best signal-background separation is achieved for the $ZZa$ SR1 analysis, 
and the worst one is obtained for the $WWa$ analysis. 

%==========================================
\section{Expected limits and comparison with existing bounds}
\label{sec:constraint}

\begin{figure*}[!ht]
  \centering  
    \includegraphics[width=0.6\textwidth]{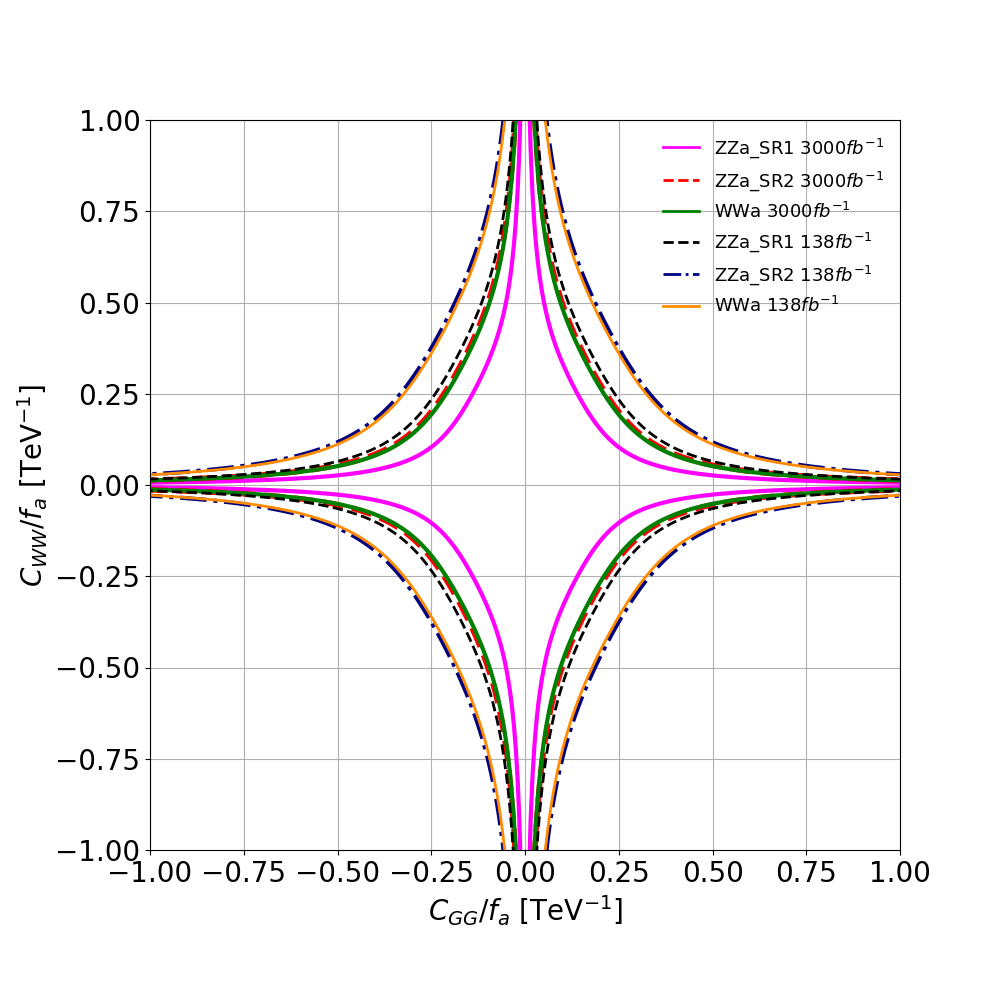}
\caption{\small Expected $95\%$ CL limits in the  $c_{WW} \mhyphen c_{GG}$  plane corresponding to  $m_a=1$ MeV obtained in the $ZZa$ SR1, $ZZa$ SR2 and $WWa$ analyses assuming an overall uncertainty of $10\%$ on the signal and background event selection efficiencies. The presented limits are based on the integrated luminosities of 138 fb$^{-1}$ (full run II of the LHC data) and  $3$ $\mathrm{ab^{-1}}$.}
\label{fig:explimits}
\end{figure*}

Using the obtained BDT response distributions, 95\% CL exclusion limits in the $c_{WW} \mhyphen c_{GG}$ plane are derived.
A threshold on the BDT output is applied to define the signal region.
The expected number of signal and background events passing this cut—computed using event weights based on cross section, luminosity, and number of generated events—are used to construct Poisson likelihood functions under the background-only and signal-plus-background hypotheses.
The CL$_{s}$ method~\cite{cl1,cl2}, implemented in the {\tt RooStats} package~\cite{Moneta:2010pm}, is used to derive the exclusion limits.
Specifically, the log-likelihood ratio test statistic $Q = 2 \ln(\mathcal{L}_{\text{signal+bkg}}/\mathcal{L}_{\text{bkg}})$ is used to compute $p$-values for both hypotheses, and the exclusion is set at 95\% confidence level by requiring
\[
\text{CL}_{s} = \frac{\mathcal{P}_{\text{signal+bkg}}(Q > Q_0)}{1 - \mathcal{P}_{\text{bkg}}(Q < Q_0)} \leq 0.05,
\]
where $Q_0$ is the observed (or expected) value of the test statistic.
Limits in the $ZZa$ (SR1 and SR2) and $WWa$ analyses are computed independently.
To account for systematic uncertainties, an overall uncertainty of 10\% is applied to both signal and background yields, reflecting detector effects such as energy scale, resolution, pile-up, and reconstruction efficiency. 
Fig.~\ref{fig:explimits} shows the expected 95\% CL limits in the $c_{WW} \mhyphen c_{GG}$ plane for the $ZZa$ SR1, $ZZa$ SR2, and $WWa$ analyses, assuming integrated luminosities of 138 fb$^{-1}$ and 3 ab$^{-1}$.
We have verified that removing the $10\%$ systematic uncertainty leads to a $5-10\%$ improvement 
in the exclusion limits across all channels, confirming the robustness of the results.

%------------------------------------------------ 
%------------------------------------------------ 

The limits have been derived under the assumption that the kinematics of the signal events 
remain independent of the values of the non-zero Wilson coefficients. 
A comparative analysis reveals that the $ZZa$ SR1 search provides slightly 
stronger constraints than both the $ZZa$ SR2 and $WWa$ analyses. 
This can be attributed to more effective background suppression and greater 
discriminating power of the BDT classifier in the fully leptonic $ZZa$ final state. 
Although the cross section for the $ZZa$ process is larger than for $WWa$, 
the key factor driving the difference in exclusion power lies in the 
distribution of the missing transverse energy ($E_T^{\text{miss}}$), 
which serves as a crucial input to the BDT. In the $ZZa$ SR1 channel, 
the dominant SM $ZZ$ background leads to very low MET, whereas in 
the $WWa$ channel, the main backgrounds such as $WW$ and $t\bar{t}$ 
produce significant MET from neutrinos, making separation from signal more challenging. 
The $ZZa$ SR2 and $WWa$ analyses yield nearly comparable limits in the $c_{WW} \mhyphen c_{GG}$ plane.

To assess the significance of our results, we compare them quantitatively with existing experimental and theoretical bounds on the ALP couplings available in the literature.
The loop-induced corrections to electroweak precision observables, encapsulated by the well-known oblique parameters \(S\), \(T\), and \(U\), were utilized in Ref.~\cite{Bauer:2017ris} to simultaneously probe the Wilson coefficients \(c_{WW} \mhyphen c_{BB}\) and \(c_{\gamma\gamma}\mhyphen c_{\gamma Z}\).  
Notably, it was found that the coefficient \(c_{BB}/f_{a}\) remains substantially unconstrained, whereas \(c_{WW}/f_{a}\) is confined to a much narrower parameter space. At the 99\% CL, an upper bound of 8~TeV\(^{-1}\) was placed on \(c_{WW}/f_{a}\).
The limits on the Wilson coefficients in the plane of \(c_{WW} \mhyphen c_{BB}\), obtained from the combination of VBS \(q_{1}q_{2} \rightarrow q'_{1}q'_{2} V_{1}V_{2}\) processes, where \(V_{1,2} = W,Z,\gamma\), are presented in Ref.~\cite{ccvv}. Assuming \(M_{V_{1}V_{2}} < 4\) TeV, values of \(c_{WW}/f_{a}\) and \(c_{BB}/f_{a}\) above 0.75~TeV\(^{-1}\) and 1.59~TeV\(^{-1}\), respectively, are excluded.
In Ref.~\cite{Hosseini:2024kuh}, the analysis was extended by imposing simultaneous two-dimensional constraints on \(c_{WW}/f_{a}\) and \(c_{a\Phi}/f_{a}\) using the measured cross sections of the \(t\bar{t}\) and \(tW\) processes at the LHC. Specifically, \(c_{a\Phi}/f_{a}\) is constrained to values around 40~TeV\(^{-1}\), while \(c_{GG}/f_{a}\) is limited to less than 4~TeV\(^{-1}\).
One-dimensional limits on \(|c_{WW}/f_a|\) for an integrated luminosity of 3~\(\mathrm{ab^{-1}}\), based on LHC searches for mono-\(W\), mono-\(Z\), \(W\gamma\), \(Z\gamma\), \(WW\), and \(WW\gamma\) final states with missing energy, are available in Refs.~\cite{Brivio:2017ije,Biswas:2023ksj,ATLAS:2018sxj,CMS:2019ppl,CMS:2023rcv,ATLAS:2019lsy,CMS:2017ret,TheATLAScollaboration:2015vog}.  
The ATLAS search for mono-\(W\) events in the leptonic final state, with an integrated luminosity of 139~\(\mathrm{fb^{-1}}\), places an upper bound of \(|c_{WW}/f_a| < 0.11\)~TeV\(^{-1}\) at 95\% CL~\cite{Biswas:2023ksj,ATLAS:2019lsy}.  
Similarly, the CMS search for mono-\(Z\) events using 35.9~\(\mathrm{fb^{-1}}\) of data provides a slightly weaker constraint of \(|c_{WW}/f_a| < 0.129\)~TeV\(^{-1}\) at 95\% CL~\cite{Biswas:2023ksj,CMS:2017nxf}.  
Additionally, a 95\% CL limit of \(|c_{WW}/f_a| < 0.44\)~TeV\(^{-1}\) can be inferred from the uncertainty on the \(Z\) boson width~\cite{Workman:2022ynf,Brivio:2017ije}.

Constraints on the ALP coupling to gluons, \(c_{GG}/f_a\), have been derived from various experimental measurements.  
The measurement of \(K^+ \to \pi^+ \nu_\ell \bar{\nu}_\ell\) provides a tighter limit of \(|c_{GG}/f_a| < 0.00769\)~TeV\(^{-1}\), while the process \(B \to K^* \nu_\ell \bar{\nu}_\ell\) yields a constraint of \(|c_{GG}/f_a| < 2.1\)~TeV\(^{-1}\)~\cite{Chakraborty:2021wda,Afik:2023mhj}.  
Mono-jet analyses at the 8~TeV LHC, using 19.6~fb\(^{-1}\) of data, result in a 95\% CL upper limit of \(|c_{GG}/f_a| < 0.025\)~TeV\(^{-1}\)~\cite{Mimasu:2014nea}.

Astrophysical observations impose stringent constraints on ALPs through their couplings to photons, electrons, and nucleons.  
For instance, stellar cooling processes set a lower bound on \(f_a\) that is several orders of magnitude stricter than those explored in this work.  
While these constraints are not directly on \(c_{GG,WW}/f_a\), they inform the allowed parameter space for ALP models in a broader context~\cite{ast1,ast2}.  
Regarding the mass range \(m_a = 1\)~MeV, supernova constraints exclude couplings in the range \(10^{-8}\) to \(10^{-5}\)~TeV\(^{-1}\)~\cite{ast3}.  
These bounds, however, are not directly comparable to our results, as they do not address the simultaneous behavior of \(c_{WW}\) and \(c_{GG}\).

%************************************************************
\section{Probe of the ALP couplings using the LHC Measurements of $W+$jets and $Z+$jets}
\label{wjzj}

\begin{figure}[ht]
	\centering
	\includegraphics[width=0.75\linewidth]{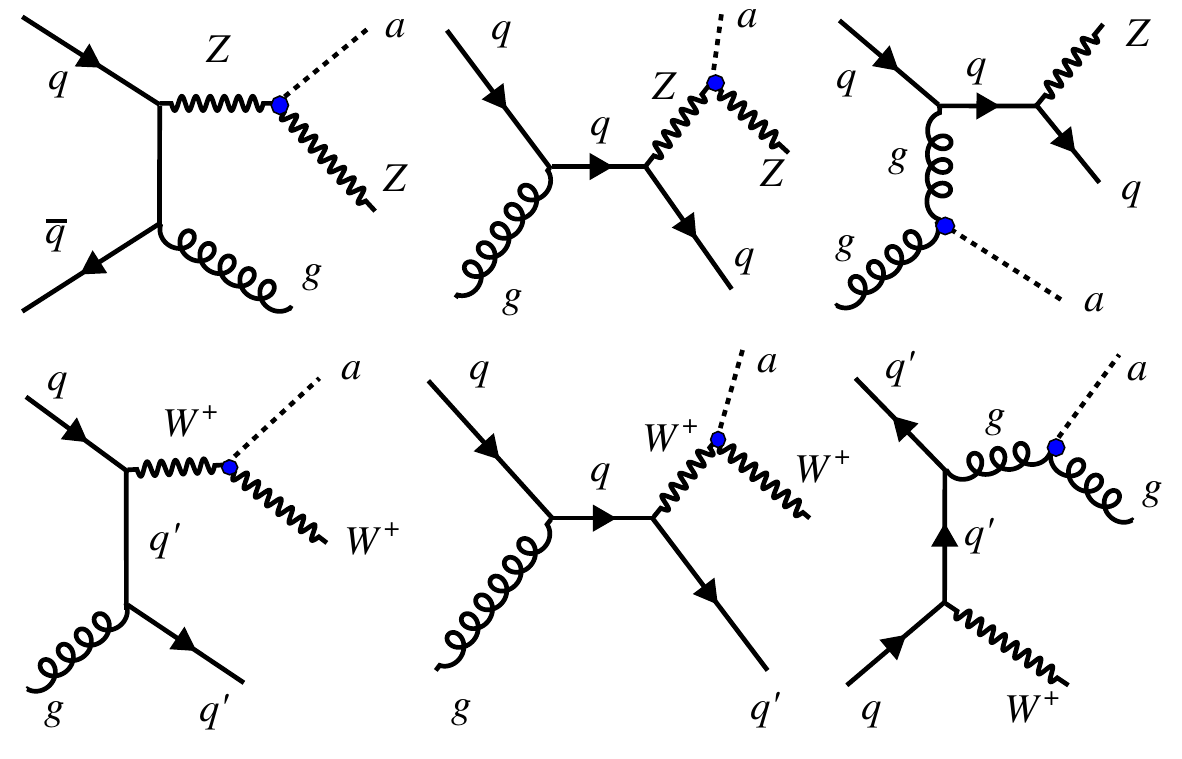}
	\caption{\small Representative leading order Feynman diagrams for production of an ALP in association with  $Z$+jets (top) and  $W$+jets (bottom) at the LHC.}
	\label{fig:vj}
\end{figure}

In this section, we analyze the sensitivity of the production of an ALP in association with a vector boson ($V$) and jets ($V$+jets+ALP), 
where $V = Z, W^{\pm}$. 
The leading-order Feynman diagrams for these processes are illustrated in Fig. \ref{fig:vj}. 
Similar to the previous analysis, we consider scenarios where the ALP does not decay within the detector, 
manifesting as missing transverse energy. 
Consequently, the final states resemble those of the SM processes $W^{\pm}$+jets and $Z$+jets, although with modified kinematics due to the presence of the ALP.

These processes are particularly sensitive to the couplings of the ALP with gluons ($c_{GG}$) and $W$ bosons ($c_{WW}$). 
In $V$+jets+ALP production, the initial state primarily involves quark-antiquark annihilation and gluon-quark interactions. 
The parton distribution function (PDF) for gluons is significantly large, particularly at low momentum fractions ($x$), 
leading to substantial contributions from gluon-induced processes, even though quark-antiquark annihilation has remarkable contribution for the $V$+jets+ALP production.

These processes are crucial as they provide a probe of both $c_{GG}$ and $c_{WW}$ couplings. 
The transverse momentum ($p_{\rm T}$) of the $V$ boson is a key observable, offering insights into the initial state kinematics and the dynamics of the associated jet production. 
A higher $p_{\rm T}$ of the $V$ boson typically corresponds to more energetic jets, reflecting the underlying parton-level interactions that characterize the final state. 
By analyzing the $p_{\rm T}$ spectrum of the $V$ boson or the leading jet, one can explore the ALP interactions with gluons and vector bosons. 
In the subsequent subsections, we utilize the $p_{\rm T}$ spectra of the $Z$ boson and leading jet, derived from CMS experiment measurements in Refs. \cite{cmszj, cmswj}, 
to probe these interactions.
The fiducial object definitions and event selection criteria were implemented manually based on the 
detailed descriptions provided in Refs.~\cite{cmszj,cmswj}, and validated by comparing the resulting 
kinematic distributions with the corresponding unfolded CMS measurements.

\subsection{$Z+$jets+ALP process}
\label{sec:zj}

In Ref.\cite{cmszj}, a measurement of the differential cross section for $Z+$jets production at a center-of-mass energy of $\sqrt{s} = 13$ TeV is presented, 
focusing on the dimuon final state $(Z \rightarrow \mu^{+}\mu^{-})$. 
The data used in this analysis were collected with the CMS detector in 2016, corresponding to an integrated luminosity of $35.9$ fb$^{-1}$. 
The study investigates regions of phase space characterized by the production of $Z$ bosons at large transverse momentum ($p_{\rm T,Z}$) 
in association with at least one high $p_{\rm T}$ jet.

In the analysis, the $Z$ boson is identified via its decay into a pair of muons ($\mu^{+}\mu^{-}$). 
Events for the $Z$+jets analysis are selected using a high-level trigger that requires a loosely isolated 
muon with a minimum $p_{\rm T}$ threshold of 24 GeV. Muon candidates are required to lie within the 
fiducial region $|\eta_{\mu}| < 2.4$ and to be separated from any jets in the event by a distance $\Delta R(\mu,\text{jets}) > 0.5$, 
where $\Delta R$ is defined as $\Delta R = \sqrt{\Delta \eta^2 + \Delta \phi^2}$. 
Two oppositely charged muons are selected, and the invariant mass of the dimuon pair, $m_{\mu^{+}\mu^{-}}$, 
is required to be consistent with the $Z$ boson mass, within the range $71 < m_{\mu^{+}\mu^{-}} < 111$ GeV.
Following the baseline selection, each muon is required to have $p_{\rm T} > 30$ GeV, and the dimuon system 
is required to have $p_{\rm T} > 200$ GeV with a rapidity $|y_{\mu^{+}\mu^{-}}| < 1.4$. 
The measured distributions are then unfolded from the detector level to the particle level. 
Jets are required to have $p_{\rm T,j} > 40$ GeV and $|\eta_{j}| < 2.4$, with at least one jet in the event required to have $p_{\rm T} > 40$ GeV.
The differential cross section is evaluated as a function of the $Z$ boson's transverse momentum, $p_{\rm T,Z}$. 
The experimental measurements of the differential $p_{\rm T,Z}$ distribution are used as input, 
along with theoretical predictions from the ALP model. A subsequent fit is performed to extract constraints on the ALP couplings $c_{GG}$ and $c_{WW}$.

\begin{figure}[ht]
	\centering
	\includegraphics[width=0.47\linewidth]{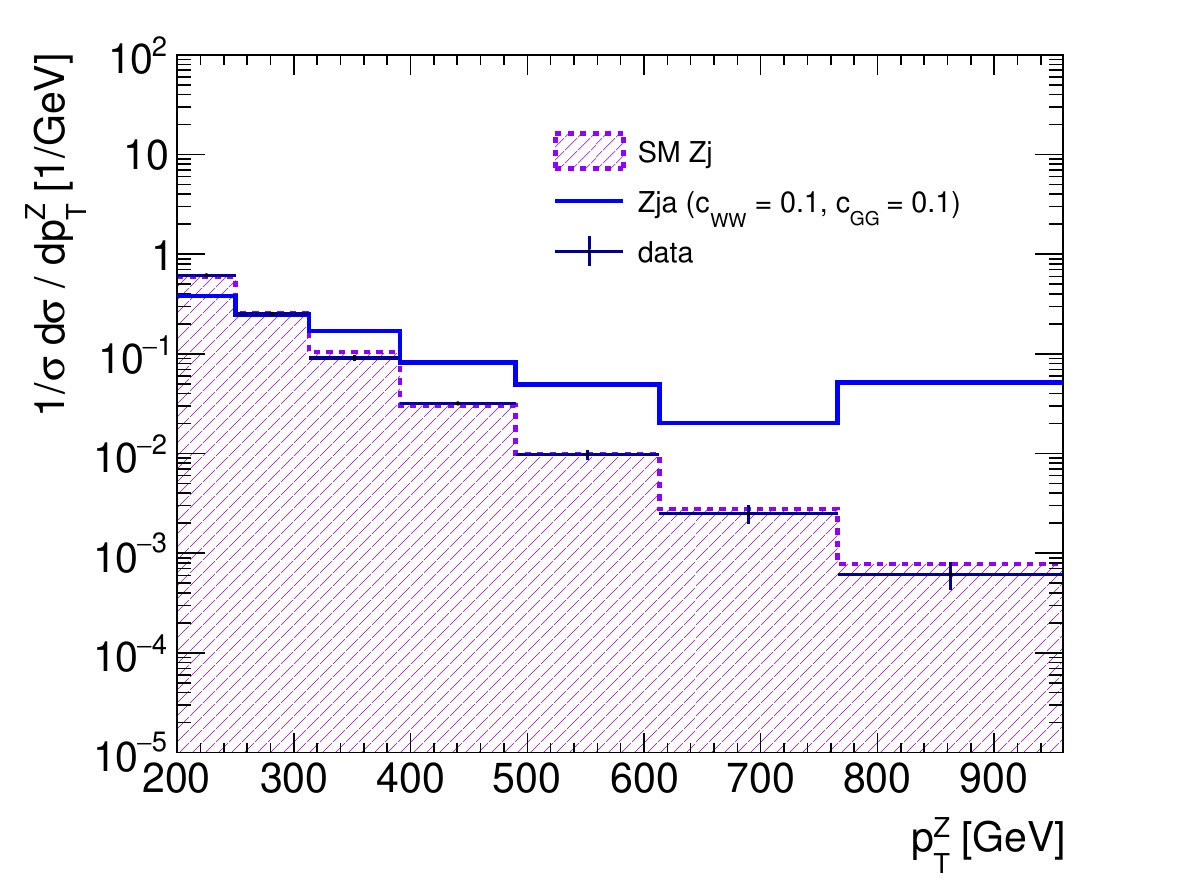}
		\includegraphics[width=0.47\linewidth]{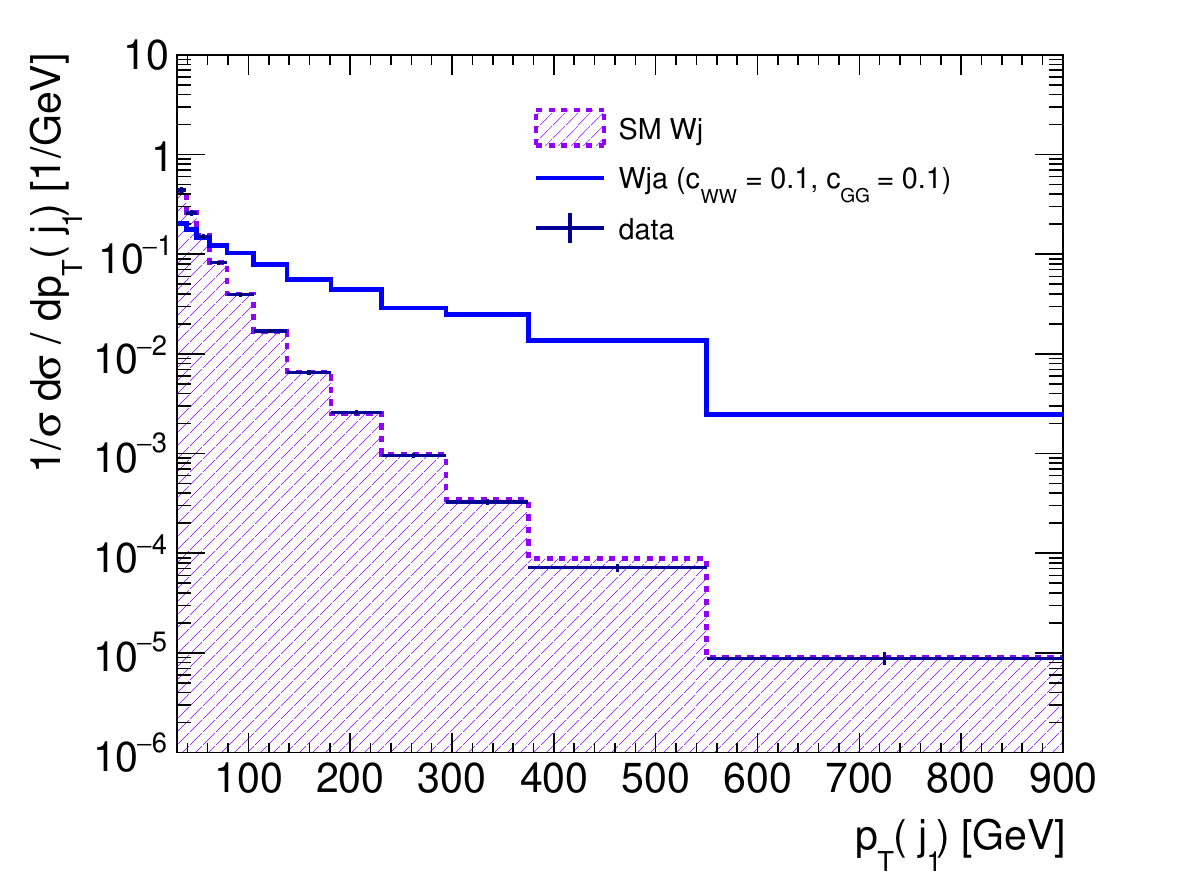}
	\caption{\small Left: The measured differential cross section as a function of the $Z$ boson $p_{\rm T}$ for $Z$+jets with the SM
	expectation and theoretical prediction of $Z$+jets+ALP where ALP escape detection and appears as missing transverse momentum.
	Right: Differential cross section measurement for the leading jet transverse momentum in the $W^{\pm}$+jets process as well as the
	SM prediction. The black circular markers represent the unfolded data measurement and the ALP expectation is displayed as solid blue line.
	For both plots the ALP mass is taken as 1 MeV.}
	\label{fig:diswj}
\end{figure}

ALP signal events are generated using \texttt{MadGraph5\_aMC@NLO}, with two of the couplings $c_{GG}$ or $c_{WW}$ 
set to a non-zero value at a time. These events are then processed through \texttt{Pythia 8} for parton showering, hadronization, 
and the decay of stable particles. After applying the selection criteria, the normalized distribution of $p_{\rm T,Z}$ is depicted in 
the left panel of Figure \ref{fig:diswj}. The distributions for ALP signal events, with $c_{GG}/f_a$ and $c_{WW}/f_a$ couplings 
each set to 0.1 TeV$^{-1}$, are also shown, assuming an ALP mass of $m_a = 1$ MeV.

To determine the upper limits on the ALP couplings, we calculate the uncorrelated 
$\chi^2$ functions of the ALP couplings and mass using the differential cross-section measurements of $p_{\rm T,Z}$. The $\chi^2$ function is defined as:

\begin{eqnarray}\label{chi2}
	\chi^2\left(\frac{c_{WW}}{f_a},\frac{c_{GG}}{f_a}, m_a\right) = \sum_{i \in \text{bins}} \frac{\left[\text{Data}(i) - \text{Bkg}(i) - \text{Pred}_{\rm ALP}(i;\frac{c_{WW}}{f_a}, \frac{c_{GG}}{f_a},m_a)\right]^2}{\delta_{i}^2},
\end{eqnarray} 

where $\text{Data}(i)$ represents the measured differential distribution in the $i$th bin, $\text{Pred}_{\rm ALP}(i)$ and $\text{Bkg}(i)$ 
represent the differential distributions of the ALP signal and the SM background, respectively, in the same bin. 
The term $\delta_i$ accounts for both systematic and statistical uncertainties. 
These uncertainties, along with the measured data distribution and the SM prediction, are derived from the 
CMS measurements of the differential cross-section for $Z$+jets production at the LHC, 
performed at a center-of-mass energy of 13 TeV with an integrated luminosity of 35.9 fb$^{-1}$ \cite{cmszj}.

The observed upper limits on the couplings $c_{GG}/f_a$ and $c_{WW}/f_a$ at a $95\%$ confidence level (CL) for an ALP mass $m_a = 1$ MeV are shown in Figure \ref{fig:res}. 
This figure also presents the projected constraints on $c_{GG}/f_a$ and $c_{WW}/f_a$ at $95\%$ CL for the High-Luminosity LHC (HL-LHC), 
assuming the systematic uncertainties remain consistent with those of the current CMS measurements.

\subsection{$W^{\pm}+$jets+ALP process}
\label{sec:wj}

In this subsection, we utilize the differential cross sections for $W$ boson production in association with jets, 
as reported in Ref.\cite{cmswj}, to explore the parameter space of the ALP model. 
These measurements were based on an integrated luminosity of 2.2 fb$^{-1}$ of data collected with the CMS detector at the LHC, 
during proton-proton collisions at a center-of-mass energy of 13 TeV.
The analysis in Ref.\cite{cmswj} measured the differential cross sections for $W$ bosons decaying via the muon channel, 
as functions of jet transverse momentum and the absolute value of rapidity for the leading jets, as well as the scalar sum of
the transverse momenta of all jets in each event ($H_{T}$). 
Additionally, the cross sections were measured as a function of the azimuthal separation between the muon direction and the leading jet.

The $W^{\pm}$+jets+ALP events were generated using \texttt{MadGraph5\_aMC@NLO}, 
followed by simulation and tuning similar to those detailed in Ref.\cite{cmswj}. 
We consider events where the ALP does not decay within the detector, 
leading to a final state that mimics the SM $W^{\pm}$+jets process but with altered kinematics due to the presence of the ALP.
The selection criteria applied are as follows: muons are required to have a transverse momentum $p_{\rm T} > 25$ GeV 
and be within the pseudorapidity range $|\eta_{\mu}| < 2.4$. Jets must satisfy the selection criteria with $p_{\rm T} > 30$ GeV and $|y_{\rm jets}| < 2.4$. 
Additionally, events are required to be in the transverse mass peak region for $W$ bosons, defined by $m^{\rm W}_{\rm T}$, 
where $m^{\rm W}_{\rm T} = \sqrt{2|\vec{p}^{~\mu}_{\rm T}| |\vec{p}^{\rm ~miss}_{T}| (1-\cos(\Delta \phi))}$. 
Here, $\Delta \phi$ is the azimuthal angle difference between the muon momentum direction and the missing transverse momentum $\vec{p}^{\rm ~miss}_{T}$.
As in the previous analysis, a fit is performed on the differential distributions to determine the sensitivity to the $c_{GG}$ and $c_{WW}$ couplings. 
Among the various differential distributions analyzed, the leading jet transverse momentum distribution provided the highest expected sensitivity. 
The differential cross-section measurement for the leading jet transverse momentum, along with the SM prediction, 
is shown in the right panel of Fig. \ref{fig:diswj}. The leading-order prediction for an ALP with $m_a = 1$ MeV is also presented.

\begin{figure}[ht]
	\centering
	\includegraphics[width=0.75\linewidth]{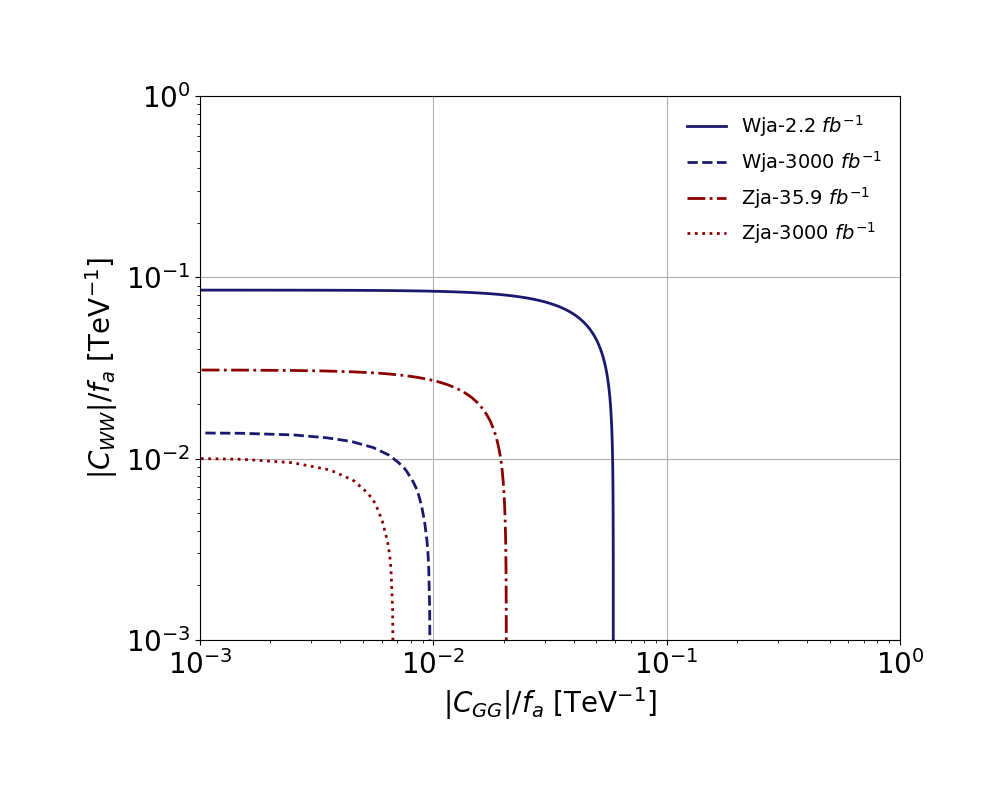}
	\caption{\small Observed $95\%$ CL limits in the $c_{WW} \mhyphen c_{GG}$ plane corresponding to $\sqrt{s}=13$ TeV and $m_a=1$ MeV obtained 
	from the measurement of the CMS experiment on $Z$+jets and $W$+jets taken from Refs.\cite{cmszj,cmswj}.  
	The projected bounds corresponding to $L_\mathrm{int.}=3$ $\mathrm{ab^{-1}}$ for the HL-LHC are displayed in dashed curves as well.}
	\label{fig:res}
\end{figure}

The upper limits on the couplings $c_{GG}/f_a$ and $c_{WW}/f_a$ at $95\%$ CL for $m_a = 1$ MeV, 
obtained from the CMS measurements of the $W$+jets process, are presented in Fig. \ref{fig:res}. 
Additionally, the projected limits for the High-Luminosity LHC (HL-LHC) are shown, 
assuming that the systematic uncertainties remain consistent with those in the current CMS measurements from Ref.\cite{cmswj}.

To thoroughly understand the results presented in Figure \ref{fig:res}, 
it is crucial to delve into the comparative sensitivity of the $Z$+jets+ALP and $W^{\pm}$+jets+ALP channels to the ALP couplings, $c_{GG}$ and $c_{WW}$.
The results clearly indicate that both channels exhibit greater sensitivity to the $c_{GG}$ coupling compared to $c_{WW}$. 
This enhanced sensitivity arises from the stronger dependence of the cross sections on $c_{GG}$. 
In the ALP model, the coupling $c_{GG}$ governs interactions between the ALP and gluons, 
which are abundant in the initial state of proton-proton collisions. 
Consequently, processes involving $c_{GG}$ tend to have larger cross sections, 
leading to a more pronounced impact on the differential distributions of observables
 like the transverse momentum of the $V$ boson (where $V = Z, W^{\pm}$) and the leading jet. 
 This results in a more significant deviation from the SM expectation when $c_{GG}$ is non-zero, 
 thereby enhancing the sensitivity of the analysis to this coupling.

When comparing the two channels under the same integrated luminosity of 3 ab$^{-1}$, the limits on the couplings are found to be relatively similar; 
however, the $Z$+jets+ALP channel consistently provides slightly better constraints than the $W^{\pm}$+jets+ALP channel. 
This difference in sensitivity can be attributed to the characteristics of the fiducial region in which the fit is performed. 
In the fiducial region for the $Z$+jets+ALP process, there is a more substantial contribution of the ALP signal relative to the SM background, 
compared to the $W^{\pm}$+jets+ALP process.  This means that the presence of the ALP has a more pronounced effect 
on the kinematic distributions in the $Z$+jets+ALP channel, allowing for tighter constraints on the ALP couplings.

In summary, while both channels offer valuable insights into the ALP couplings, the $Z$+jets+ALP channel stands out as slightly more sensitive. 
This enhanced sensitivity is mostly due to  more significant contribution of the ALP signal in the fiducial region where the analysis is conducted.

%***********************************************************
\section{Summary and conclusions} 
\label{sumconvvv}
Axion-like particles appear in theories with a spontaneously broken global $U(1)$ symmetry. 
ALPs provide the possibility to address some of the long-lasting SM problems, e.g. the dark matter problem, 
baryon asymmetry problem, neutrino mass problem. Motivated by such a potential, 
we examined the capability of the 14 TeV HL-LHC to explore the parameter space of an ALP with a mass of 1 MeV. 
Such an ALP predominantly decays into a pair of photons. However, due to the small mass and the strong experimental 
limits on the ALP-photon coupling, the ALP decay width into photons is so small that it is likely to be stable at the detector 
manifesting itself as missing energy. Hence, the ALP production process features a relatively large missing energy. 
The assumed production processes are the production of an ALP in association with a pair of electroweak gauge 
bosons, i.e. $pp\rightarrow ZZa$ and $pp\rightarrow WWa$. These processes are sensitive to the Wilson 
coefficients $c_{WW} $ and $c_{GG}$ , and are used to set expected $95\%$ CL exclusion limits in the $c_{WW} \mhyphen c_{GG}$  plane. 
The integrated luminosities assumed to calculate the limits  are 138 fb$^{-1}$ corresponding to the data collected at Run 2 of the LHC and
 3 $\mathrm{ab}^{-1}$, which is the ultimate integrated luminosity planned for the HL-LHC. 
 The signal regions SR1 and SR2 which respectively correspond to the fully leptonic ($ZZ\rightarrow 4\ell$) and semi-leptonic ($ZZ\rightarrow 2\ell 2j$) 
 final states are considered in the $ZZa$ analysis. The $WWa$ search is focused on the fully leptonic ($WW\rightarrow 2\ell2\nu$) final state. 
 Signal and dominant SM backgrounds events are generated with a realistic simulation of the detector response. 
 With the help of a proper set of discriminating variables, the ALP production processes are separated from the SM background. 
 To achieve the best signal-background discrimination, a multivariate technique is deployed. We presented the expected 
 two-dimensional exclusion limits obtained in these analyses. Comparing the expected limits obtained in this study with 
 the limits derived from the LHC searches, namely the searches for the mono-$W$ and mono-$Z$  plus missing energy signals, 
 shows that a significant region of the parameter space which has been unprobed becomes accessible with the help of the present study. 
The presented results correspond to the ALP mass of 1 MeV. However, it can be seen that for lower (sub-MeV) ALP masses,
 the limits would not change significantly. 
 It can be concluded that the present analysis can serve as a tool for searching for the associated production of
  an ALP and a pair of electroweak gauge bosons at the HL-LHC, and can be used to probe the ALP couplings to $W$ bosons, 
  gluons  to an unprecedented extent in certain regions of the parameter space.

In the final part of this paper, we have thoroughly investigated the sensitivity of  ALP production 
in association with a massive vector boson and jets, $V$+jets+ALP, where $V = Z, W^{\pm}$. 
We focused on scenarios where the ALP does not decay within the detector, resulting in missing transverse energy, 
leading to final states that mimic the SM processes $W^{\pm}$+jets and $Z$+jets, but with distinct kinematic signatures due to the ALP presence. 
We utilized differential cross-section measurements from the CMS experiment for $W^{\pm}$+jets and $Z$+jets processes to constrain the ALP 
parameter space, considering both $c_{WW}$ and $c_{GG}$ couplings simultaneously. 
Our analysis demonstrates that fitting these differential distributions enables probing of ALP couplings 
down to $10^{-3}$ TeV$^{-1}$ at the High-Luminosity LHC (HL-LHC).  The $Z$+jets+ALP channel exhibits superior sensitivity with respect to $W^{\pm}$+jets, 
primarily due to the more contribution of the ALP signal in the fiducial region of the analysis.
In addition to the relatively larger ALP signal contribution in the fiducial region, the $Z$+jets+ALP channel 
benefits from several experimental advantages that contribute to its superior sensitivity. 
The CMS $Z$+jets measurement used in this analysis is based on 35.9 fb$^{-1}$ of data, 
compared to only 2.2 fb$^{-1}$ for the $W$+jets measurement. 
Moreover, being a more recent result, the $Z$+jets analysis benefits from improved detector 
understanding and reduced systematic uncertainties. 
The key observable in the $Z$+jets channel, the transverse momentum of the $Z$ boson, 
is reconstructed from muon kinematics, which inherently have smaller experimental uncertainties 
than jet-based or MET-related observables. 
These factors collectively enhance the precision and exclusion power of the $Z$+jets+ALP analysis.

The two channels studied-$VV+a$ and $V$+jets-offer complementary sensitivity to ALP couplings, with the $V$+jets analysis 
yielding stronger constraints due to its enhanced kinematic reach and larger production rates. 
Importantly, both analyses constrain the ALP parameter space simultaneously in the $(c_{WW}, c_{GG})$ plane, and their 
combination could further tighten these bounds. 
The consistent interpretation across both channels highlights the robustness of the limits obtained and underscores 
the value of multi-channel approaches in probing ALP interactions at the LHC.

%***********************************************************
\section*{Acknowledgments}
The authors would like to thank Gh. Haghighat for his valuable assistance 
in the preparation of portions of the text, as well as for the insightful discussions and constructive comments.

%***********************************************************
\RaggedRight % removes extra spaces between words in bibliography


\begin{thebibliography}{}

%\cite{Peccei:1977hh}
\bibitem{Peccei:1977hh} 
  R.~D.~Peccei and H.~R.~Quinn,
  %``CP Conservation in the Presence of Instantons,''
  Phys.\ Rev.\ Lett.\  {\bf 38}, 1440 (1977).
  doi:10.1103/PhysRevLett.38.1440
  %%CITATION = doi:10.1103/PhysRevLett.38.1440;%%
  %4580 citations counted in INSPIRE as of 16 Dec 2018

%\cite{Hook:2018dlk}
\bibitem{Hook:2018dlk}
A.~Hook,
%``TASI Lectures on the Strong CP Problem and Axions,''
PoS \textbf{TASI2018}, 004 (2019)
[arXiv:1812.02669 [hep-ph]].
%159 citations counted in INSPIRE as of 19 Sep 2024

%\cite{Dine:2000cj}
\bibitem{Dine:2000cj}
M.~Dine,
%``TASI lectures on the strong CP problem,''
[arXiv:hep-ph/0011376 [hep-ph]].
%126 citations counted in INSPIRE as of 19 Sep 2024

%\cite{DiLuzio:2020wdo}
\bibitem{DiLuzio:2020wdo}
L.~Di Luzio, M.~Giannotti, E.~Nardi and L.~Visinelli,
%``The landscape of QCD axion models,''
[arXiv:2003.01100 [hep-ph]].
%14 citations counted in INSPIRE as of 27 May 2020

%\cite{Hook:2019qoh}
\bibitem{Hook:2019qoh}
A.~Hook, S.~Kumar, Z.~Liu and R.~Sundrum,
%``The High Quality QCD Axion and the LHC,''
[arXiv:1911.12364 [hep-ph]].
%2 citations counted in INSPIRE as of 27 May 2020

%\cite{Quevillon:2019zrd} 
\bibitem{Quevillon:2019zrd}
J.~Quevillon and C.~Smith,
%``Axions are blind to anomalies,''
Eur. Phys. J. C \textbf{79}, no.10, 822 (2019)
doi:10.1140/epjc/s10052-019-7304-4
[arXiv:1903.12559 [hep-ph]].
%4 citations counted in INSPIRE as of 27 May 2020

 %\cite{Bellazzini:2017neg}
\bibitem{Bellazzini:2017neg}
B.~Bellazzini, A.~Mariotti, D.~Redigolo, F.~Sala and J.~Serra,
%``$R$-axion at colliders,''
Phys. Rev. Lett. \textbf{119}, no.14, 141804 (2017)
doi:10.1103/PhysRevLett.119.141804
[arXiv:1702.02152 [hep-ph]].
%22 citations counted in INSPIRE as of 12 Jun 2020

%\cite{Arganda:2018cuz}    
\bibitem{Arganda:2018cuz}
E.~Arganda, A.~D.~Medina, N.~I.~Mileo, R.~A.~Morales and A.~Szynkman,
%``Constraining R-axion models through dijet searches at the LHC,''
Phys. Lett. B \textbf{789}, 575-581 (2019)
doi:10.1016/j.physletb.2018.12.064
[arXiv:1808.01292 [hep-ph]].
%3 citations counted in INSPIRE as of 12 Jun 2020

%\cite{Dine:1981rt}
\bibitem{Dine:1981rt}
M.~Dine, W.~Fischler and M.~Srednicki,
%``A Simple Solution to the Strong CP Problem with a Harmless Axion,''
Phys. Lett. B \textbf{104}, 199-202 (1981)
doi:10.1016/0370-2693(81)90590-6
%3553 citations counted in INSPIRE as of 24 Aug 2024

%\cite{Preskill:1982cy}
\bibitem{Preskill:1982cy}
J.~Preskill, M.~B.~Wise and F.~Wilczek,
%``Cosmology of the Invisible Axion,''
Phys. Lett. B \textbf{120}, 127-132 (1983)
doi:10.1016/0370-2693(83)90637-8
%3482 citations counted in INSPIRE as of 19 Sep 2024

%\cite{Abbott:1982af}
\bibitem{Abbott:1982af}
L.~Abbott and P.~Sikivie,
%``A Cosmological Bound on the Invisible Axion,''
Phys. Lett. B \textbf{120}, 133-136 (1983)
doi:10.1016/0370-2693(83)90638-X
%1807 citations counted in INSPIRE as of 12 Jun 2020

%\cite{Dine:1982ah}  
\bibitem{Dine:1982ah}
M.~Dine and W.~Fischler,
%``The Not So Harmless Axion,''
Phys. Lett. B \textbf{120}, 137-141 (1983)
doi:10.1016/0370-2693(83)90639-1
%1769 citations counted in INSPIRE as of 12 Jun 2020


%\cite{Jeong:2018jqe}
\bibitem{Jeong:2018jqe}
K.~S.~Jeong, T.~H.~Jung and C.~S.~Shin,
%``Adiabatic electroweak baryogenesis driven by an axionlike particle,''
Phys. Rev. D \textbf{101}, no.3, 035009 (2020)
doi:10.1103/PhysRevD.101.035009
[arXiv:1811.03294 [hep-ph]].
%32 citations counted in INSPIRE as of 19 Sep 2024
  
 %\cite{Co:2019wyp}
\bibitem{Co:2019wyp}
R.~T.~Co and K.~Harigaya,
%``Axiogenesis,''
Phys. Rev. Lett. \textbf{124}, no.11, 111602 (2020)
doi:10.1103/PhysRevLett.124.111602
[arXiv:1910.02080 [hep-ph]].
%7 citations counted in INSPIRE as of 27 May 2020

%\cite{Bauer:2019gfk}
\bibitem{Bauer:2019gfk}
M.~Bauer, M.~Neubert, S.~Renner, M.~Schnubel and A.~Thamm,
%``Axionlike Particles, Lepton-Flavor Violation, and a New Explanation of $a_\mu$ and $a_e$,''
Phys. Rev. Lett. \textbf{124}, no.21, 211803 (2020)
doi:10.1103/PhysRevLett.124.211803
[arXiv:1908.00008 [hep-ph]].
%57 citations counted in INSPIRE as of 25 Apr 2021

%\cite{Cornella:2019uxs}
\bibitem{Cornella:2019uxs}
C.~Cornella, P.~Paradisi and O.~Sumensari,
%``Hunting for ALPs with Lepton Flavor Violation,''
JHEP \textbf{01}, 158 (2020)
doi:10.1007/JHEP01(2020)158
[arXiv:1911.06279 [hep-ph]].
%40 citations counted in INSPIRE as of 25 Apr 2021

%\cite{Krasznahorkay:2015iga}
\bibitem{Krasznahorkay:2015iga}
A.~J.~Krasznahorkay, M.~Csatl\'os, L.~Csige, Z.~G\'acsi, J.~Guly\'as, M.~Hunyadi, T.~J.~Ketel, A.~Krasznahorkay, I.~Kuti and B.~M.~Nyak\'o, \textit{et al.}
%``Observation of Anomalous Internal Pair Creation in Be8 : A Possible Indication of a Light, Neutral Boson,''
Phys. Rev. Lett. \textbf{116}, no.4, 042501 (2016)
doi:10.1103/PhysRevLett.116.042501
[arXiv:1504.01527 [nucl-ex]].
%379 citations counted in INSPIRE as of 19 Sep 2024 

%\cite{Ellwanger:2016wfe}
\bibitem{Ellwanger:2016wfe}
U.~Ellwanger and S.~Moretti,
%``Possible Explanation of the Electron Positron Anomaly at 17 MeV in $^8Be$ Transitions Through a Light Pseudoscalar,''
JHEP \textbf{11}, 039 (2016)
doi:10.1007/JHEP11(2016)039
[arXiv:1609.01669 [hep-ph]].
%97 citations counted in INSPIRE as of 19 Sep 2024

 
%\cite{Kitahara:2019lws}
\bibitem{Kitahara:2019lws}
T.~Kitahara, T.~Okui, G.~Perez, Y.~Soreq and K.~Tobioka,
%``New physics implications of recent search for $K_L \to \pi^0 \nu\bar{\nu}$ at KOTO,''
Phys. Rev. Lett. \textbf{124}, no.7, 071801 (2020)
doi:10.1103/PhysRevLett.124.071801
[arXiv:1909.11111 [hep-ph]].
%26 citations counted in INSPIRE as of 26 May 2020

 %\cite{Dias:2014osa}
\bibitem{Dias:2014osa}
A.~Dias, A.~Machado, C.~Nishi, A.~Ringwald and P.~Vaudrevange,
%``The Quest for an Intermediate-Scale Accidental Axion and Further ALPs,''
JHEP \textbf{06}, 037 (2014)
doi:10.1007/JHEP06(2014)037
[arXiv:1403.5760 [hep-ph]].
%126 citations counted in INSPIRE as of 06 May 2020

%\cite{Chen:2012baa}
\bibitem{Chen:2012baa}
C.~S.~Chen and L.~H.~Tsai,
%``Peccei-Quinn symmetry as the origin of Dirac Neutrino Masses,''
Phys. Rev. D \textbf{88}, no.5, 055015 (2013)
doi:10.1103/PhysRevD.88.055015
[arXiv:1210.6264 [hep-ph]].
%16 citations counted in INSPIRE as of 06 May 2020

%\cite{Salvio:2015cja}
\bibitem{Salvio:2015cja}
A.~Salvio,
%``A Simple Motivated Completion of the Standard Model below the Planck Scale: Axions and Right-Handed Neutrinos,''
Phys. Lett. B \textbf{743}, 428-434 (2015)
doi:10.1016/j.physletb.2015.03.015
[arXiv:1501.03781 [hep-ph]].
%44 citations counted in INSPIRE as of 06 May 2020

%\cite{Brivio:2017ije}
\bibitem{Brivio:2017ije} 
  I.~Brivio, M.~B.~Gavela, L.~Merlo, K.~Mimasu, J.~M.~No, R.~del Rey and V.~Sanz,
  %``ALPs Effective Field Theory and Collider Signatures,''
  Eur.\ Phys.\ J.\ C {\bf 77}, no. 8, 572 (2017)
  doi:10.1140/epjc/s10052-017-5111-3
  [arXiv:1701.05379 [hep-ph]].
  %%CITATION = doi:10.1140/epjc/s10052-017-5111-3;%%
  %31 citations counted in INSPIRE as of 16 Dec 2018


%\cite{Biswas:2023ksj}
\bibitem{Biswas:2023ksj}
T.~Biswas,
%``Probing the interactions of axion-like particles with electroweak bosons and the Higgs boson in the high energy regime at LHC,''
JHEP \textbf{05}, 081 (2024)
doi:10.1007/JHEP05(2024)081
[arXiv:2312.05992 [hep-ph]].
%5 citations counted in INSPIRE as of 15 Sep 2024


\bibitem{Choi:2020rgn}
K.~Choi, S.~H.~Im and C.~Sub Shin,
%``Recent Progress in the Physics of Axions and Axion-Like Particles,''
Ann. Rev. Nucl. Part. Sci. \textbf{71} (2021), 225-252
doi:10.1146/annurev-nucl-120720-031147
[arXiv:2012.05029 [hep-ph]].

\bibitem{Athron:2020maw}
P.~Athron, C.~Bal\'azs, A.~Beniwal, J.~E.~Camargo-Molina, A.~Fowlie, T.~E.~Gonzalo, S.~Hoof, F.~Kahlhoefer, D.~J.~E.~Marsh and M.~T.~Prim, \textit{et al.}
%``Global fits of axion-like particles to XENON1T and astrophysical data,''
JHEP \textbf{05} (2021), 159
doi:10.1007/JHEP05(2021)159
[arXiv:2007.05517 [astro-ph.CO]].

\bibitem{Han:2020dwo}
C.~Han, M.~L.~L\'opez-Ib\'a\~nez, A.~Melis, O.~Vives and J.~M.~Yang,
``Anomaly-free leptophilic axionlike particle and its flavor violating tests,''
Phys. Rev. D \textbf{103} (2021) no.3, 035028
doi:10.1103/PhysRevD.103.035028
[arXiv:2007.08834 [hep-ph]].

%\cite{Mimasu:2014nea}
\bibitem{Mimasu:2014nea} 
  K.~Mimasu and V.~Sanz,
  %``ALPs at Colliders,''
  JHEP {\bf 1506}, 173 (2015)
  doi:10.1007/JHEP06(2015)173
  [arXiv:1409.4792 [hep-ph]].
  %%CITATION = doi:10.1007/JHEP06(2015)173;%%
  %40 citations counted in INSPIRE as of 16 Dec 2018  	
  
%\cite{Bauer:2017ris}
\bibitem{Bauer:2017ris} 
  M.~Bauer, M.~Neubert and A.~Thamm,
  %``Collider Probes of Axion-Like Particles,''
  JHEP {\bf 1712}, 044 (2017)
  doi:10.1007/JHEP12(2017)044
  [arXiv:1708.00443 [hep-ph]].
  %%CITATION = doi:10.1007/JHEP12(2017)044;%%
  %39 citations counted in INSPIRE as of 16 Dec 2018

%\cite{Aloni:2019ruo}
\bibitem{Aloni:2019ruo}
D.~Aloni, C.~Fanelli, Y.~Soreq and M.~Williams,
%``Photoproduction of Axionlike Particles,''
Phys. Rev. Lett. \textbf{123}, no.7, 071801 (2019)
doi:10.1103/PhysRevLett.123.071801
[arXiv:1903.03586 [hep-ph]].
%18 citations counted in INSPIRE as of 27 May 2020
  
%\cite{Haghighat:2021djz}
\bibitem{Haghighat:2021djz}
G.~Haghighat and M.~Mohammadi Najafabadi,
%``Search for lepton-flavor-violating ALPs at a future muon collider and utilization of polarization-induced effects,''
Nucl. Phys. B \textbf{980}, 115827 (2022)
doi:10.1016/j.nuclphysb.2022.115827
[arXiv:2106.00505 [hep-ph]].
%17 citations counted in INSPIRE as of 12 Feb 2024

%\cite{Haghighat:2020nuh}
\bibitem{Haghighat:2020nuh}
G.~Haghighat, D.~Haji Raissi and M.~Mohammadi Najafabadi,
%``New collider searches for axionlike particles coupling to gluons,''
Phys. Rev. D \textbf{102}, no.11, 115010 (2020)
doi:10.1103/PhysRevD.102.115010
[arXiv:2006.05302 [hep-ph]].
%13 citations counted in INSPIRE as of 12 Feb 2024

%\cite{Haghighat:2022qyh}
\bibitem{Haghighat:2022qyh}
G.~Haghighat, M.~Mohammadi Najafabadi, K.~Sakurai and W.~Yin,
%``Probing a light dark sector at future lepton colliders via invisible decays of the SM-like and dark Higgs bosons,''
Phys. Rev. D \textbf{107}, no.3, 035033 (2023)
doi:10.1103/PhysRevD.107.035033
[arXiv:2209.07565 [hep-ph]].
%6 citations counted in INSPIRE as of 12 Feb 2024


  %\cite{Ebadi:2019gij}
\bibitem{Ebadi:2019gij}
J.~Ebadi, S.~Khatibi and M.~Mohammadi Najafabadi,
%``New probes for axionlike particles at hadron colliders,''
Phys. Rev. D \textbf{100}, no.1, 015016 (2019)
doi:10.1103/PhysRevD.100.015016
[arXiv:1901.03061 [hep-ph]].
%8 citations counted in INSPIRE as of 27 May 2020

%\cite{Hosseini:2024kuh}
\bibitem{Hosseini:2024kuh}
Y.~Hosseini and M.~Mohammadi Najafabadi,
%``Exploring axionlike particle couplings through single top tW-channel and top pair production at the LHC,''
Phys. Rev. D \textbf{110}, no.5, 055026 (2024)
doi:10.1103/PhysRevD.110.055026
[arXiv:2408.11588 [hep-ph]].
%0 citations counted in INSPIRE as of 18 Sep 2024


%\cite{Hosseini:2022tac}
\bibitem{Hosseini:2022tac}
Y.~Hosseini and M.~Mohammadi Najafabadi,
%``Prospects for Probing Axionlike Particles at a Future Hadron Collider through Top Quark Production,''
Universe \textbf{8}, no.6, 301 (2022)
doi:10.3390/universe8060301
[arXiv:2208.00414 [hep-ph]].
%3 citations counted in INSPIRE as of 24 Aug 2024


%\cite{ARGUS:1995bjh}
\bibitem{ARGUSlimit}
H.~Albrecht \textit{et al.} [ARGUS],
%``A Search for lepton flavor violating decays tau ----\ensuremath{>} e alpha, tau ---\ensuremath{>} mu alpha,''
Z. Phys. C \textbf{68}, 25-28 (1995)
doi:10.1007/BF01579801
%81 citations counted in INSPIRE as of 19 Sep 2024



%\cite{Baldini:2020okg}
\bibitem{Baldini:2020okg}
A.~M.~Baldini \textit{et al.} [MEG],
%``Search for lepton flavour violating muon decay mediated by a new light particle in the MEG experiment,''
Eur. Phys. J. C \textbf{80}, no.9, 858 (2020)
doi:10.1140/epjc/s10052-020-8364-1
[arXiv:2005.00339 [hep-ex]].
%4 citations counted in INSPIRE as of 01 May 2021

%\cite{Aad:2019ugc}
\bibitem{Aad:2019ugc}
G.~Aad \textit{et al.} [ATLAS],
%``Searches for lepton-flavour-violating decays of the Higgs boson in $\sqrt{s}=13$ TeV pp collisions with the ATLAS detector,''
Phys. Lett. B \textbf{800}, 135069 (2020)
doi:10.1016/j.physletb.2019.135069
[arXiv:1907.06131 [hep-ex]].
%28 citations counted in INSPIRE as of 01 May 2021

%\cite{Calibbi:2020jvd}
\bibitem{Calibbi:2020jvd}
L.~Calibbi, D.~Redigolo, R.~Ziegler and J.~Zupan,
%``Looking forward to lepton-flavor-violating ALPs,''
JHEP \textbf{09}, 173 (2021)
doi:10.1007/JHEP09(2021)173
[arXiv:2006.04795 [hep-ph]].
%150 citations counted in INSPIRE as of 19 Sep 2024

%\cite{Iguro:2020rby}
\bibitem{Iguro:2020rby}
S.~Iguro, Y.~Omura and M.~Takeuchi,
%``Probing $\mu\tau$ flavor-violating solutions for the muon $g-2$ anomaly at Belle II,''
JHEP \textbf{09}, 144 (2020)
doi:10.1007/JHEP09(2020)144
[arXiv:2002.12728 [hep-ph]].
%7 citations counted in INSPIRE as of 03 Jun 2021

%\cite{Endo:2020mev}
\bibitem{Endo:2020mev}
M.~Endo, S.~Iguro and T.~Kitahara,
%``Probing $e\mu$ flavor-violating ALP at Belle II,''
JHEP \textbf{06}, 040 (2020)
doi:10.1007/JHEP06(2020)040
[arXiv:2002.05948 [hep-ph]].
%19 citations counted in INSPIRE as of 03 Jun 2021


%\cite{Cepeda:2025diq}
\bibitem{c144}
M.~Cepeda, J.~M.~No, C.~Ramos, R.~M.~Sand\'a Seoane and J.~Zurita,
%``Exotic $h \to Z a$ Higgs decays into $\tau$ leptons,''
[arXiv:2503.08781 [hep-ph]].
%0 citations counted in INSPIRE as of 09 Jun 2025

%\cite{Liu:2023bby}
\bibitem{c244}
J.~Liu, Y.~Luo and M.~Song,
%``Investigation of the concurrent effects of ALP-photon and ALP-electron couplings in Collider and Beam Dump Searches,''
JHEP \textbf{09}, 104 (2023)
doi:10.1007/JHEP09(2023)104
[arXiv:2304.05435 [hep-ph]].
%20 citations counted in INSPIRE as of 09 Jun 2025

%\cite{Bisht:2024hbs}
\bibitem{c344}
D.~Bisht, S.~Chakraborty and A.~Samanta,
%``A comprehensive study of ALPs from $B$-decays,''
[arXiv:2412.09678 [hep-ph]].
%3 citations counted in INSPIRE as of 09 Jun 2025



%\cite{Bauer:2017nlg}
\bibitem{Bauer:2017nlg}
M.~Bauer, M.~Neubert and A.~Thamm,
%``LHC as an Axion Factory: Probing an Axion Explanation for $(g-2)_\mu$ with Exotic Higgs Decays,''
Phys. Rev. Lett. \textbf{119}, no.3, 031802 (2017)
doi:10.1103/PhysRevLett.119.031802
[arXiv:1704.08207 [hep-ph]].
%82 citations counted in INSPIRE as of 19 Sep 2024
  
%\cite{Chatrchyan:2012cg}
\bibitem{Chatrchyan:2012cg}
S.~Chatrchyan \textit{et al.} [CMS],
%``Search for a Non-Standard-Model Higgs Boson Decaying to a Pair of New Light Bosons in Four-Muon Final States,''
Phys. Lett. B \textbf{726}, 564-586 (2013)
doi:10.1016/j.physletb.2013.09.009
[arXiv:1210.7619 [hep-ex]].
%64 citations counted in INSPIRE as of 07 May 2020 

%\cite{Khachatryan:2017mnf}
\bibitem{Khachatryan:2017mnf}
V.~Khachatryan \textit{et al.} [CMS],
%``Search for light bosons in decays of the 125 GeV Higgs boson in proton-proton collisions at $ \sqrt{s}=8 $ TeV,''
JHEP \textbf{10}, 076 (2017)
doi:10.1007/JHEP10(2017)076
[arXiv:1701.02032 [hep-ex]].
%72 citations counted in INSPIRE as of 07 May 2020

%\cite{Jaeckel:2015jla}
\bibitem{Jaeckel:2015jla}
J.~Jaeckel and M.~Spannowsky,
%``Probing MeV to 90 GeV axion-like particles with LEP and LHC,''
Phys. Lett. B \textbf{753}, 482-487 (2016)
doi:10.1016/j.physletb.2015.12.037
[arXiv:1509.00476 [hep-ph]].
%277 citations counted in INSPIRE as of 19 Sep 2024

 %\cite{Knapen:2016moh}
\bibitem{Knapen:2016moh}
S.~Knapen, T.~Lin, H.~K.~Lou and T.~Melia,
%``Searching for Axionlike Particles with Ultraperipheral Heavy-Ion Collisions,''
Phys. Rev. Lett. \textbf{118}, no.17, 171801 (2017)
doi:10.1103/PhysRevLett.118.171801
[arXiv:1607.06083 [hep-ph]].
%67 citations counted in INSPIRE as of 07 May 2020

%\cite{Inoue:2008zp}
\bibitem{Inoue:2008zp}
Y.~Inoue, Y.~Akimoto, R.~Ohta, T.~Mizumoto, A.~Yamamoto and M.~Minowa,
%``Search for solar axions with mass around 1 eV using coherent conversion of axions into photons,''
Phys. Lett. B \textbf{668}, 93-97 (2008)
doi:10.1016/j.physletb.2008.08.020
[arXiv:0806.2230 [astro-ph]].
%123 citations counted in INSPIRE as of 06 May 2020 

%\cite{Arik:2008mq}
\bibitem{Arik:2008mq}
E.~Arik \textit{et al.} [CAST],
%``Probing eV-scale axions with CAST,''
JCAP \textbf{02}, 008 (2009)
doi:10.1088/1475-7516/2009/02/008
[arXiv:0810.4482 [hep-ex]].
%241 citations counted in INSPIRE as of 06 May 2020


\bibitem{Irastorza:2013dav}
I.~Irastorza \textit{et al.} [IAXO],
%``The International Axion Observatory IAXO. Letter of Intent to the CERN SPS committee,''
CERN-SPSC-2013-022.
%68 citations counted in INSPIRE as of 19 Sep 2024


%\cite{Dobrich:2019dxc}
\bibitem{Dobrich:2019dxc}
B.~Döbrich, J.~Jaeckel and T.~Spadaro,
%``Light in the beam dump. Axion-Like Particle production from decay photons in proton beam-dumps,''
JHEP \textbf{05}, 213 (2019)
doi:10.1007/JHEP05(2019)213
[arXiv:1904.02091 [hep-ph]].
%21 citations counted in INSPIRE as of 26 May 2020

%\cite{ATLAS:2018sxj}
\bibitem{ATLAS:2018sxj}
M.~Aaboud \textit{et al.} [ATLAS],
%``Search for heavy resonances decaying to a photon and a hadronically decaying $Z/W/H$ boson in $pp$ collisions at $\sqrt{s}=13$ $\mathrm{TeV}$ with the ATLAS detector,''
Phys. Rev. D \textbf{98}, no.3, 032015 (2018)
doi:10.1103/PhysRevD.98.032015
[arXiv:1805.01908 [hep-ex]].
%45 citations counted in INSPIRE as of 27 Mar 2024

%\cite{CMS:2019ppl}    
\bibitem{CMS:2019ppl}
A.~M.~Sirunyan \textit{et al.} [CMS],
%``Search for anomalous triple gauge couplings in WW and WZ production in lepton + jet events in proton-proton collisions at $\sqrt{s} =$ 13 TeV,''
JHEP \textbf{12}, 062 (2019)
doi:10.1007/JHEP12(2019)062
[arXiv:1907.08354 [hep-ex]].
%52 citations counted in INSPIRE as of 27 Mar 2024

%\cite{CMS:2023rcv}
\bibitem{CMS:2023rcv}
A.~Hayrapetyan \textit{et al.} [CMS],
%``Observation of WW$\gamma$ production and search for H$\gamma$ production in proton-proton collisions at $\sqrt{s}$ = 13 TeV,''
Phys. Rev. Lett. \textbf{132}, no.12, 121901 (2024)
doi:10.1103/PhysRevLett.132.121901
[arXiv:2310.05164 [hep-ex]].
%6 citations counted in INSPIRE as of 27 Mar 2024

%\cite{ATLAS:2019lsy}
\bibitem{ATLAS:2019lsy}
G.~Aad \textit{et al.} [ATLAS],
%``Search for a heavy charged boson in events with a charged lepton and missing transverse momentum from $pp$ collisions at $\sqrt{s} = 13$ TeV with the ATLAS detector,''
Phys. Rev. D \textbf{100}, no.5, 052013 (2019)
doi:10.1103/PhysRevD.100.052013
[arXiv:1906.05609 [hep-ex]].
%143 citations counted in INSPIRE as of 27 Mar 2024

%\cite{CMS:2017ret}
\bibitem{CMS:2017ret}
A.~M.~Sirunyan \textit{et al.} [CMS],
%``Search for dark matter and unparticles in events with a Z boson and missing transverse momentum in proton-proton collisions at $ \sqrt{s}=13 $ TeV,''
JHEP \textbf{03}, 061 (2017)
[erratum: JHEP \textbf{09}, 106 (2017)]
doi:10.1007/JHEP03(2017)061
[arXiv:1701.02042 [hep-ex]].
%62 citations counted in INSPIRE as of 27 Mar 2024

%\cite{TheATLAScollaboration:2015vog}
\bibitem{TheATLAScollaboration:2015vog}
%``Search for new resonances in events with one lepton and missing transverse momentum in $pp$ collisions at $\sqrt s = 13$ TeV with the ATLAS detector,''
ATLAS-CONF-2015-063.
%12 citations counted in INSPIRE as of 27 Mar 2024

%\cite{CMS:2017nxf}
\bibitem{CMS:2017nxf}
A.~M.~Sirunyan \textit{et al.} [CMS],
%``Search for new physics in events with a leptonically decaying Z boson and a large transverse momentum imbalance in proton\textendash{}proton collisions at $\sqrt{s} $ = 13 $\,\text {TeV}$,''
Eur. Phys. J. C \textbf{78}, no.4, 291 (2018)
doi:10.1140/epjc/s10052-018-5740-1
[arXiv:1711.00431 [hep-ex]].
%80 citations counted in INSPIRE as of 27 Mar 2024




%\cite{ATLAS:2023ian}
\bibitem{c1}
G.~Aad \textit{et al.} [ATLAS],
%``Search for short- and long-lived axion-like particles in $H\rightarrow a a \rightarrow 4\gamma $ decays with the ATLAS experiment at the LHC,''
Eur. Phys. J. C \textbf{84}, no.7, 742 (2024)
doi:10.1140/epjc/s10052-024-12979-0
[arXiv:2312.03306 [hep-ex]].
%19 citations counted in INSPIRE as of 21 Apr 2025

%\cite{ATLAS:2020hii}
\bibitem{c2}
G.~Aad \textit{et al.} [ATLAS],
%``Measurement of light-by-light scattering and search for axion-like particles with 2.2 nb$^{-1}$ of Pb+Pb data with the ATLAS detector,''
JHEP \textbf{03}, 243 (2021)
[erratum: JHEP \textbf{11}, 050 (2021)]
doi:10.1007/JHEP03(2021)243
[arXiv:2008.05355 [hep-ex]].
%173 citations counted in INSPIRE as of 21 Apr 2025


%\cite{CMS:2021xor}
\bibitem{c3}
A.~Tumasyan \textit{et al.} [CMS],
%``Search for heavy resonances decaying to ZZ or ZW and axion-like particles mediating nonresonant ZZ or ZH production at $ \sqrt{s} $ = 13 TeV,''
JHEP \textbf{04}, 087 (2022)
doi:10.1007/JHEP04(2022)087
[arXiv:2111.13669 [hep-ex]].
%36 citations counted in INSPIRE as of 21 Apr 2025


%\cite{ZurbanoFernandez:2020cco}
\bibitem{ZurbanoFernandez:2020cco}
I.~Zurbano Fernandez, M.~Zobov, A.~Zlobin, F.~Zimmermann, M.~Zerlauth, C.~Zanoni, C.~Zannini, O.~Zagorodnova, I.~Zacharov and M.~Yu, \textit{et al.}
%``High-Luminosity Large Hadron Collider (HL-LHC): Technical design report,''
CERN, 2020,
ISBN 978-92-9083-586-8, 978-92-9083-587-5
doi:10.23731/CYRM-2020-0010
%347 citations counted in INSPIRE as of 27 Mar 2024


%\cite{Bonilla:2022pxu}
\bibitem{ccvv}
J.~Bonilla, I.~Brivio, J.~Machado-Rodr\'\i{}guez and J.~F.~de Troc\'oniz,
%``Nonresonant searches for axion-like particles in vector boson scattering processes at the LHC,''
JHEP \textbf{06}, 113 (2022)
doi:10.1007/JHEP06(2022)113
[arXiv:2202.03450 [hep-ph]].
%33 citations counted in INSPIRE as of 16 Sep 2024


%\cite{Alloul:2013bka}
\bibitem{Alloul:2013bka} 
  A.~Alloul, N.~D.~Christensen, C.~Degrande, C.~Duhr and B.~Fuks,
  %``FeynRules  2.0 - A complete toolbox for tree-level phenomenology,''
  Comput.\ Phys.\ Commun.\  {\bf 185}, 2250 (2014)
  doi:10.1016/j.cpc.2014.04.012
  [arXiv:1310.1921 [hep-ph]].
  %%CITATION = doi:10.1016/j.cpc.2014.04.012;%%
  %1021 citations counted in INSPIRE as of 16 Dec 2018

%\cite{Degrande:2011ua}
\bibitem{Degrande:2011ua}
C.~Degrande, C.~Duhr, B.~Fuks, D.~Grellscheid, O.~Mattelaer and T.~Reiter,
%``UFO - The Universal FeynRules Output,''
Comput. Phys. Commun. \textbf{183}, 1201-1214 (2012)
doi:10.1016/j.cpc.2012.01.022
[arXiv:1108.2040 [hep-ph]].
%1364 citations counted in INSPIRE as of 19 Sep 2024
  
  
%\cite{Alwall:2011uj}
\bibitem{Alwall:2011uj} 
  J.~Alwall, M.~Herquet, F.~Maltoni, O.~Mattelaer and T.~Stelzer,
  %``MadGraph 5 : Going Beyond,''
  JHEP {\bf 1106}, 128 (2011)
  doi:10.1007/JHEP06(2011)128
  [arXiv:1106.0522 [hep-ph]].
  %%CITATION = doi:10.1007/JHEP06(2011)128;%%
  %2764 citations counted in INSPIRE as of 16 Dec 2018

%\cite{Ball:2012cx}
\bibitem{Ball:2012cx}
R.~D.~Ball, V.~Bertone, S.~Carrazza, C.~S.~Deans, L.~Del Debbio, S.~Forte, A.~Guffanti, N.~P.~Hartland, J.~I.~Latorre and J.~Rojo, \textit{et al.}
%``Parton distributions with LHC data,''
Nucl. Phys. B \textbf{867}, 244-289 (2013)
doi:10.1016/j.nuclphysb.2012.10.003
[arXiv:1207.1303 [hep-ph]].
%2791 citations counted in INSPIRE as of 19 Sep 2024

%\cite{deFavereau:2013fsa}
\bibitem{deFavereau:2013fsa} 
  J.~de Favereau {\it et al.} [DELPHES 3 Collaboration],
  %``DELPHES 3, A modular framework for fast simulation of a generic collider experiment,''
  JHEP {\bf 1402}, 057 (2014)
  doi:10.1007/JHEP02(2014)057
  [arXiv:1307.6346 [hep-ex]].
  %%CITATION = doi:10.1007/JHEP02(2014)057;%%
  %1117 citations counted in INSPIRE as of 16 Dec 2018
  
  %\cite{GEANT4:2002zbu}
\bibitem{geant}
S.~Agostinelli \textit{et al.} [GEANT4],
%``GEANT4 - A Simulation Toolkit,''
Nucl. Instrum. Meth. A \textbf{506}, 250-303 (2003)
doi:10.1016/S0168-9002(03)01368-8
%20179 citations counted in INSPIRE as of 25 Apr 2025
    
%\cite{Cacciari:2008gp}
\bibitem{Cacciari:2008gp} 
  M.~Cacciari, G.~P.~Salam and G.~Soyez,
  %``The anti-$k_t$ jet clustering algorithm,''
  JHEP {\bf 0804}, 063 (2008)
  doi:10.1088/1126-6708/2008/04/063
  [arXiv:0802.1189 [hep-ph]].
  %%CITATION = doi:10.1088/1126-6708/2008/04/063;%%
  %5635 citations counted in INSPIRE as of 16 Dec 2018
  
  %\cite{Cacciari:2011ma}
\bibitem{Cacciari:2011ma} 
  M.~Cacciari, G.~P.~Salam and G.~Soyez,
  %``FastJet User Manual,''
  Eur.\ Phys.\ J.\ C {\bf 72}, 1896 (2012)
  doi:10.1140/epjc/s10052-012-1896-2
  [arXiv:1111.6097 [hep-ph]].
  %%CITATION = doi:10.1140/epjc/s10052-012-1896-2;%%
  %2646 citations counted in INSPIRE as of 16 Dec 2018
  
\bibitem{Hocker:2007ht} 
A.~Hocker {\it et al.},
%``TMVA - Toolkit for Multivariate Data Analysis,''
PoS ACAT {\bf }, 040 (2007)
[physics/0703039 [PHYSICS]].

%\cite{Speckmayer:2010zz}
\bibitem{Speckmayer:2010zz}
P.~Speckmayer, A.~Hocker, J.~Stelzer and H.~Voss,
%``The toolkit for multivariate data analysis, TMVA 4,''
J. Phys. Conf. Ser. \textbf{219}, 032057 (2010)
doi:10.1088/1742-6596/219/3/032057
%147 citations counted in INSPIRE as of 16 Feb 2024

%\cite{Therhaag:2010zz}
\bibitem{Therhaag:2010zz}
J.~Therhaag,
%``TMVA Toolkit for multivariate data analysis in ROOT,''
PoS \textbf{ICHEP2010}, 510 (2010)
doi:10.22323/1.120.0510
%20 citations counted in INSPIRE as of 16 Feb 2024


%\cite{Xia:2018cfz}
\bibitem{Xia:2018cfz}
L.~G.~Xia,
%``Understanding the boosted decision tree methods with the weak-learner approximation,''
[arXiv:1811.04822 [physics.data-an]].
%7 citations counted in INSPIRE as of 25 Apr 2025

%\cite{Junk:1999kv}
\bibitem{cl1}
T.~Junk,
%``Confidence level computation for combining searches with small statistics,''
Nucl. Instrum. Meth. A \textbf{434}, 435-443 (1999)
doi:10.1016/S0168-9002(99)00498-2
[arXiv:hep-ex/9902006 [hep-ex]].
%2345 citations counted in INSPIRE as of 25 Sep 2022

%\cite{Read:2002hq}
\bibitem{cl2}
A.~L.~Read,
%``Presentation of search results: The CL(s) technique,''
J. Phys. G \textbf{28}, 2693-2704 (2002)
doi:10.1088/0954-3899/28/10/313
%3797 citations counted in INSPIRE as of 25 Sep 2022

%\cite{Moneta:2010pm}
\bibitem{Moneta:2010pm}ZZa
L.~Moneta, K.~Belasco, K.~S.~Cranmer, S.~Kreiss, A.~LZZaaro, D.~Piparo, G.~Schott, W.~Verkerke and M.~Wolf,
%``The RooStats Project,''
PoS \textbf{ACAT2010}, 057 (2010)
doi:10.22323/1.093.0057
[arXiv:1009.1003 [physics.data-an]].
%752 citations counted in INSPIRE as of 11 Apr 2023

%\cite{ParticleDataGroup:2022pth}
\bibitem{Workman:2022ynf}
R.~L.~Workman \textit{et al.} [Particle Data Group],
%``Review of Particle Physics,''
PTEP \textbf{2022}, 083C01 (2022)
doi:10.1093/ptep/ptac097
%4852 citations counted in INSPIRE as of 25 Apr 2025




%\cite{Chakraborty:2021wda}
\bibitem{Chakraborty:2021wda}
S.~Chakraborty, M.~Kraus, V.~Loladze, T.~Okui and K.~Tobioka,
%``Heavy QCD axion in b\textrightarrow{}s transition: Enhanced limits and projections,''
Phys. Rev. D \textbf{104}, no.5, 055036 (2021)
[erratum: Phys. Rev. D \textbf{108}, no.3, 039903 (2023)]
doi:10.1103/PhysRevD.104.055036
[arXiv:2102.04474 [hep-ph]].
%50 citations counted in INSPIRE as of 16 Sep 2024


%\cite{Afik:2023mhj}
\bibitem{Afik:2023mhj}
Y.~Afik, B.~D\"obrich, J.~Jerhot, Y.~Soreq and K.~Tobioka,
%``Probing long-lived axions at the KOTO experiment,''
Phys. Rev. D \textbf{108}, no.5, 5 (2023)
doi:10.1103/PhysRevD.108.055007
[arXiv:2303.01521 [hep-ph]].
%11 citations counted in INSPIRE as of 16 Sep 2024

\bibitem{ast1}
%\cite{Cadamuro:2011fd}
D.~Cadamuro and J.~Redondo,
%``Cosmological bounds on pseudo Nambu-Goldstone bosons,''
JCAP \textbf{02}, 032 (2012)
doi:10.1088/1475-7516/2012/02/032
[arXiv:1110.2895 [hep-ph]].
%297 citations counted in INSPIRE as of 04 Jan 2025

\bibitem{ast2}
%\cite{Raffelt:1996wa}
G.~G.~Raffelt,
%``Stars as laboratories for fundamental physics: The astrophysics of neutrinos, axions, and other weakly interacting particles,''
1996, ISBN 978-0-226-70272-8
%161 citations counted in INSPIRE as of 04 Jan 2025

\bibitem{ast3}
%\cite{Fukuda:2015ana}
H.~Fukuda, K.~Harigaya, M.~Ibe and T.~T.~Yanagida,
%``Model of visible QCD axion,''
Phys. Rev. D \textbf{92}, no.1, 015021 (2015)
doi:10.1103/PhysRevD.92.015021
[arXiv:1504.06084 [hep-ph]].
%138 citations counted in INSPIRE as of 04 Jan 2025


%\cite{CMS:2021fxy}
\bibitem{cmszj}
A.~M.~Sirunyan \textit{et al.} [CMS],
%``Measurements of the differential cross sections of the production of Z + jets and $\gamma$ + jets and of Z boson emission collinear with a jet in pp collisions at $ \sqrt{s} $ = 13 TeV,''
JHEP \textbf{05}, 285 (2021)
doi:10.1007/JHEP05(2021)285
[arXiv:2102.02238 [hep-ex]].
%11 citations counted in INSPIRE as of 19 Aug 2024

%\cite{CMS:2017gbl}
\bibitem{cmswj}
A.~M.~Sirunyan \textit{et al.} [CMS],
%``Measurement of the differential cross sections for the associated production of a $W$ boson and jets in proton-proton collisions at $\sqrt{s}=13$ TeV,''
Phys. Rev. D \textbf{96}, no.7, 072005 (2017)
doi:10.1103/PhysRevD.96.072005
[arXiv:1707.05979 [hep-ex]].
%79 citations counted in INSPIRE as of 19 Aug 2024





\end{thebibliography}
\end{document}